# Image encryption using fractional integral transforms: Vulnerabilities, threats and future scope

Gurpreet Kaur[1] , Rekha Agarwal[1] and Vinod Patidar[2,*]

[1]Amity Institute of Information Technology, Amity University, Noida 201303, India
[2] Sir Padampat Singhania University, Bhatewar, Udaipur 313 601, Rajasthan, India

## Abstract

With the enormous usage of digital media in almost every sphere from education to entertainment, the security of sensitive information has been a concern. As images are the most frequently used means to convey information, therefore the issue related to the privacy preservation needs to be addressed in each of the application domains. There are various security methods proposed by researchers from time to time. This paper presents a review of various image encryption schemes based on fractional integral transform. As the fractional integral transforms have evolved through their applications from optical signal processing to digital signal and digital image processing over the decades. In this article, we have adopted an architecture and corresponding domain-based taxonomy to classify various existing schemes in the literature. The schemes are classified according to the implementation platform, that may be an optical setup comprising of the Spatial modulators, lenses and charged coupled devices or it can be a mathematical modelling of such transforms. Various schemes are classified according to the methodology adopted in each of them and a comparative analysis is also presented in tabular form.  Based on the observations, the work is converged into a summary of various challenges and some constructive guidelines are provided for consideration in future works. Such a narrative review of encryption algorithm based on various architectural schematics in fractional integral transforms has not been presented before at one place.

**\*Corresponding Author**
 vinod.patidar@spsu.ac.in; patidar.vinod@gmail.com

# 1. Introduction

Fractional transforms are the generalization of full transforms which we refer to as ordinary transforms in a more generic sense. Interestingly, the idea of fractional order in a transform first came into existence in 1695 during discussions between Leibnez and L' Hospital [1]: "Can the meaning of derivatives with integer order be generalized to derivatives with non-integer orders?" The question that was put up more than 300 years ago did not get a solution till the work on fractional calculus got explored. Later Jean- Baptiste Joseph Fourier in 1807 made important contributions to the study of trigonometric series and claimed that a periodic signal could be represented by a series of harmonically related sinusoids for the solution of 1D problems. Thus, the well-known Fourier transform is named in honour of Joseph Fourier for his significant contribution and application of the Fourier transform (FT) in many scientific disciplines. However, with the ever-expanding scope of research, it was found that FT has some shortcomings. As it is a holistic transform, the time domain signal is converted to the frequency domain and therefore is able to analyse only time-invariant signals. In other words, it is not possible to obtain a local time-frequency analysis which is pivotal for processing a time-variant or nonstationary signal. Thus, fractional Fourier transforms (FrFT), Short time Fourier transform (STFT), Wigner- Ville distribution, Wavelet transform, Gabor transform etc. were proposed as an alternative.

The initial work on fractional transform by Namias [2] presented a theory on fractional powers of Fourier transform and its application to quantum mechanics. The formal mathematical elaboration to Namias's theory was given by Mc Bride and Kerr [3]. Later, Lohmann [4] illustrated the relation of FrFT to Wigner rotation and image rotation. Almeida [5] further elaborated the concept by proposing a time-frequency representation of FrFT. Further, Ozakatas and Mendelovic proposed optical implementation and interpretation of FrFT [6, 7, 8]. With the evolution of digital channels, the digital computation of FrFT [9], and its discrete version [10] gave a new perspective to the application of FrFT in optical signal processing and related applications [11]. Pei et al [12] established a relationship between FrFT and DFrFT (Discrete fractional Fourier transform) using Hermite eigen vectors based on the postulate in [13]. Various methods of DFrFT representations are given [14, 15, 16] with the extension to other similar transform domains [17, 18, 19, 20]. We won't elaborate much on the mathematical details of the transforms here, interested readers may refer to above-mentioned references for the mathematical aspect of integral transforms and more specifically fractional Fourier transform and its variants. However, we give a conceptual description of the definition of fractional integral transforms. The term 'fractional' in a transform indicates that some parameter has non-integer value. We can define any integral transform of the input function, $f(x)$ using any transform operator, T as:

$$\mathrm{T}[f(x)](u) = \int_{-\infty}^{\infty} K(x,u)f(x)dx \qquad (1)$$

where $K(x,u)$ is operator kernel. For example, in Fourier transform, $K(x,u) = \exp(-i2\pi ux)$. If it is a fractional transform then the operator is denoted as $T^{\alpha}$ with 'α' as a parameter of fractionalization. Therefore,

$$\mathrm{T}^{\alpha}[f(x)](u) = \int_{-\infty}^{\infty} K(\alpha,x,u)f(x)dx \qquad (2)$$

For instance, continuous fractional Fourier transform is the generalization of a continuous Fourier transform. The a-th order continuous fractional Fourier Transform of a function, $y(t)$ is given as:

$$Y_{\alpha}(u) = \int_{-\infty}^{+\infty} Q_a(u,t)y(t)dt \qquad (2.a)$$

where $Q_a(u,t)$ is transform kernel given by

$$Q_a(u,t) = \sqrt{1-jcot\alpha} \cdot e^{j\pi(t^2 cot\alpha - 2tucsc(\alpha) + u^2 cot\alpha)}$$

$$= \sum_{k=0}^{\infty} exp\left(-\frac{jk\alpha\pi}{2}\right)\psi_k(t) \cdot \psi_k(u) \qquad (2.b)$$

$\psi_k(t)$ is $kth$ order Hermite Gaussian function, $\alpha = a\pi/2$

$$\psi_k(t) = \frac{2^{\frac{1}{4}}}{\sqrt{2^k\, k!}}\, H_k\big(\sqrt{2\pi t}\big) e^{-\pi t^2} \qquad (2.c)$$

where $H_k$ is $k^{th}$ Hermite polynomial with $k$ real zeros.

For the discrete version of these fractional transforms, the postulate of discrete Fourier transform (DFT) is followed. As, $N \times N$ DFT matrix $F$ is defined as

$$F_{kn} = \frac{1}{\sqrt{N}}\, e^{-\frac{j2\pi}{N}\cdot kn} \quad 0 \le k, n \le N-1 \qquad (2.d)$$

where $N$ is the length of the input sequence. Thus, $\alpha th$ order $N \times N$ discrete fractional Fourier transform (DFRFT) matrix is defined [12] as :

$$F^\alpha = V\, \Lambda^a\, V^T \qquad (2.e)$$

$$= \begin{cases} \sum_{k=0}^{N-1} e^{-\frac{j\pi}{2}ka} v_k v_k^T, & \text{for } N:\text{odd} \\ \sum_{k=0}^{N-2} e^{-\frac{j\pi}{2}ka} v_k v_k^T + e^{-\frac{j\pi}{2}Na} v_N v_N^T, & \text{for } N:\text{even} \end{cases} \qquad (2.f)$$

where $V = [v_1\, v_2\, \ldots\, v_{N-2}\, v_{N-1}]$ for $N:odd$ and $V = [v_1\, v_2\, \ldots\, v_{N-2}\, v_N]$ for $N:even$, $v_k$ is kth-order Hermite-gaussian like eigenvector, $\Lambda$ is diagonal matrix with its diagonal entries corresponding to eigenvalues of each column vector $v_k$. However, there are certain properties [6] [7] [2] that are desirable for fractional integral transform used in Eq. (2). Some of them are:

1. The fractional transform has to be continuous for any real value of the parameter, 'α'.
2. It should be additive: $T^{\alpha_1+\alpha_2} = T^{\alpha_1} \cdot T^{\alpha_2}$.
3. It should be reproducible for full transform if the parameter is replaced by integer values.
4. For $\alpha = 1$, it should give $T^1 = T$, a full transform.
5. For $\alpha = 0$, it should give $T^0 = I$, an identity matrix.
6. From the additivity property,

$$\int_{-\infty}^{\infty} K(\alpha_1, x, u) \cdot K(\alpha_2, y, u)\, du = K(\alpha_1 + \alpha_2, x, y) \qquad (3)$$

It is likely to mention here that the fractional parameter in a fractional Fourier transform refers to an angle of rotation (Wigner distribution) [4]. In some references, the fractional parameter is represented as $\alpha = a\pi/2$, where $a$: fractional number. If the angle of rotation, $\alpha = 0$, the transform is said to be in purely time domain. If $\alpha = 1$, it gives the transformation to the frequency domain whereas if the parameter is some fractional value then the transformation output results in a collective time-frequency domain. Table 1 lists some of the fractional transforms that are used in various applications of signal processing. Very few of them are used for image encryption applications due to certain properties that are required to be fulfilled for cryptographic applications.

**Table 1** Various fractional integral transforms

| Frequently used | Less frequently used |
|---|---|
| Fractional Fourier Transforms [15,47,49,51,89,99,100,102,103,104, 106, 118, 134, 143, 146, 148,149,151,152, 153,156,157,161, 170, 193, 194, 195] | Fractional Riesz Transforms |
| Fractional Cosine Transform [ 18, 20, 105, 108, 157, 171] | Fractional F-Kravchuk Transform |
| Fractional Sine Transforms [18, 20] | Fractional Cauchy Transforms |
| Fractional Hartley Transforms [52, 75, 115, 116, 117] | Fractional Slant Transform |
| Fractional Mellin Transforms [71, 96, 97, 98, 110] | Fractional Stieltjes Transforms |
| Fractional Angular Transform [112, 113, 137, 154, 169] | Fractional Abel Transforms |
| Fractional Hadamard Transforms [19] | Fractional Sumudu Transforms |
| Fractional Gyrator Transform [43, 50, 54, 95, 129, 164, 174] | Fractional Brownian Transforms |
| Fractional Hilbert Transforms | Fractional Walsh Transforms |
| Fractional Affine Transforms | Fractional JigsawTransforms |
| Fractional Random Transforms | Fractional Kekre Transforms |
| Fractional Hankel Transforms | Fractional Schrodinger Transforms |
| Fractional Radon Transforms | Fractional Riemann Derivative |
| Fractional Wigner Distribution | Fractional Fokker-Plank Equation |
| Fractional DCT Transforms | Fractional Lagendre Transform |
| Fractional Hilbert Transforms | |
| Fractional Laplace Transforms | |
| Fractional S -Transform | |
| Fractional Wavelet Transforms [129,131] | |
| Fractional Dual Tree Complex Wavelet Transform | |
| Fractional Haar Transforms | |
| Fractional Polar Harmonic Transform | |

## *1.1 Contributions and outline*

The major contributions of this review article are summarized as:

- Information regarding the background and evolution of fractional integral transforms and their application in image encryption.
- Detailed a taxonomy on various methods and corresponding architectural schematics for implementing these transforms in different domains.
- A brief overview and recent developments in optical transforms for image encryption with a tabulated description of recent review articles and various cryptanalytic strategies that are adopted to break the encryption.
- Review recent articles on the digital implementation of fractional integral transforms that have been merged with other domains/schemes for enhanced of security. Each of the classification is separately described and reviewed.
- The performance parameters adopted to evaluate an image encryption scheme are also summarized for reference in the comparative analysis of schemes.
- Based on the observations made in the review article, some issues are highlighted along with some viable solutions. A set of constructive guidelines are summarized that may be helpful to future researchers in designing a robust and highly sensitive encryption algorithm based on digital implementation of these fractional integral transforms.

The paper is further organized into five more sections. Section 2 provides the taxonomy along with a description of each classification and the review. Section 3 elaborates on the performance measures of encryption algorithms. Section 4 provides a comparative analysis of the results of some recently proposed articles. A summary on observations based on the literature review is included in Section 5. The review is concluded in Section 6.

## 2. Taxonomy of fractional integral transforms

The fractional integral transforms have evolved through their applications from optical signal processing to digital signal and digital image processing over the decades. In this article, we have adopted an architecture and corresponding domain-based taxonomy to classify various existing schemes in the literature. The architecture can be broadly classified on the bases of the platforms used for implementation as shown in Fig. 1. The platform can be an optical setup that comprises of lenses, spatial light modulators (SLM), and charge-coupled devices (CCD). Another platform is based on the use of random phase masks (RPM) in transforming image pixels. Yet another is a digital platform, where mathematical modelling is followed to achieve the transformation.

## *2.1 Optical data processing*

Optical data processing got introduced almost four decades before by Van der Lugt as an optical correlator which is based on the usage of the thin lens to produce two dimensional Fourier transform of an image. This further led to the invention of other more advanced optical and optoelectronic processors. The classical methods for the realization of the optical scheme are based on two architectures [21]: a 4f-Vander Lugt (VL) and a joint transform correlator (JTC) architecture. In both of these methods, the input image is displayed in the form of transparency or as on spatial light modulator (SLM). With the advancement in technology, SLMs that are used these days are electrically addressed liquid crystal-based SLMs. The randomness in phase is obtained with ground glass or with a nonuniform coating of gelatine on glass plates. The random phase masks (RPMs) thus obtained are recorded on SLMs during encryption or decryption. The outcome of a DRPE encryption is a random noise-like pattern with complex nature. In order to record these complex coefficients for storage and transmission, a holographic technique is required. Although both architectures, require two RPMs to convert an image (amplitude or phase) to a stationary random noise, JTC is considered superior to VLC architecture. The VLC architecture requires conjugate RPMs and stringent alignment for decryption whereas JTC does not require these two conditions and it is considered as alleviated from these limitations. Hence a JTC architecture is considered superior to the VLC. To record the decrypted image either a CCD (charged couple device) or a conjugate of input plane RPM is used. In another method known as the optical phase conjugation method [22], a conjugation of an encrypted image is obtained with the use of optical phase conjugation in a photo refractive crystal through 4 wave mixing. This phase conjugation can nullify the effect of RPM in the decryption process. A most recent classical implementation of fractional Fourier transform in terms of wave functions is presented in [23].

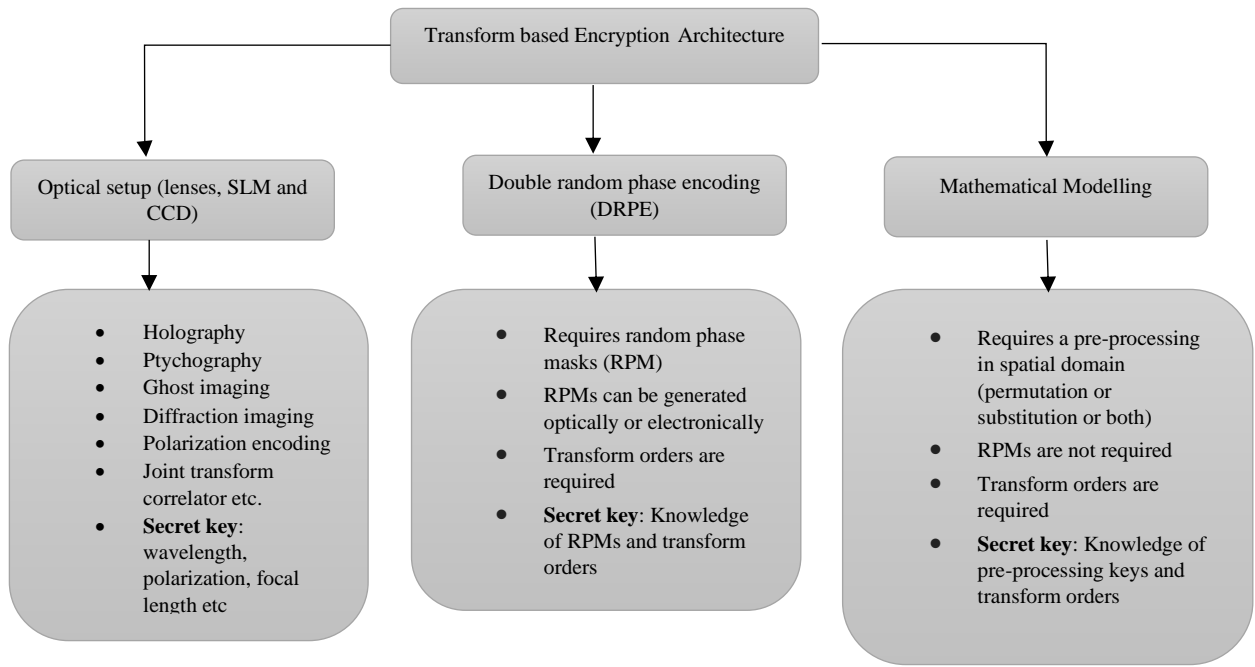

**Fig. 1.** Classification of architectures for fractional transform-based image encryption

We provide a brief overview of the various optical setups that are used for obtaining an optical transform of the scene or image. These are categorized as:

- **Holographic methods**: Holography is based on using an interference pattern generated by diffraction of the light field in 3 dimensions. Their resultant 3D image retains depth, parallax and other such properties of the scene. Thus, the hologram is an unintelligible pattern formed by an image. Digital holography is further divided into two categories viz. Off-axis digital Holography and Phase-shifting digital holography. Javidi et al. [24] first presented a combined approach to providing image security through DRPE (Double Random Phase encryption) and holography. The author further extended his work to 3D information encryption [25]. Some of the most recent reviews are available in the literature [26] [27] that give insight into the evolution of this scheme over the last decade.
- **Ptychography:** It is based on coherent imaging generated by using multiple probes that generate multiple diffraction patterns in a far field. Ptychography offers good quality of both, recovered amplitude and phase distribution. Similar to holography, it also generates complex amplitude of the object but it does not require any reference beam like in holography. The application of Ptychography in image encryption has been proposed by many researchers [28] [29] [30] and most recently in [31] [32].
- **Ghost imaging**: It is also known as coherent imaging or two-photon imaging or photon correlated imaging. It's a technique that produces an image formed by combining effects from two light detectors: one from the multipixel detector that does not view the object and another is a single pixel detector that views the object. Clemente et al. [33] proposed to use of ghost imaging for image encryption. Some of the recent works [34] [35] are based on a similar strategy.
- **Diffractive imaging**: It is referred to as imaging formed by a highly coherent beam of wavelike particles like electrons, x-rays or other wavelike particles. The waves thus diffracted from the object form a pattern which is recorded on a detector. The pattern is used to reconstruct an image with an iterative feedback algorithm. The advantage of the absence of lenses is that the final image has no aberrations and therefore resolution is only dependent on the wavelength, aperture size and exposure. The application of diffractive imaging in image encryption is proposed in [36] [37] [38]
- **Polarization encoding**: An optical plane wave is used to illuminate the intensity key image and encoded into a polarization state. It is then passed through a polarizer (pixelated polarizer) to obtain the encrypted image. Gopinathan et al. [39] proposed to use of polarization encoding in image encryption. Some of the recent works in encryption application is proposed in [40].
- **Joint Transform Correlators**: The joint power spectrum of the plane image and key codes are the encrypted data in the joint transform correlators [41]. Joint correlator-based encryption uses the same key code for decryption as used in encryption. This is unlike a classical DRPE scheme where a conjugate key is required. Many encryption schemes have been recently proposed based on joint transform correlator in fractional transform domain [42] [43].
- **Phase retrieval method**: In addition to the methods described above, there is an iterative phase retrieval method [44] [45] [46] wherein a digital approach is usually applied for embedding the input image into POM (phase only mask) and either a digital or optical method is employed for image decryption. The main objective of a phase retrieval algorithm is to find either the correct or an estimate of POM under some constraint for a measured amplitude. Phase retrieval algorithms can be 2D or 3D. Unlike holographic-based or diffractive imaging-based optical encoding, a phase retrieval-based optical security system generates phase-only masks as ciphertexts. Various transform domains such as FrFT and Gyrator transform can be employed in these encoding schemes.

*Advantages of optical encryption*
1. Optical instruments such as SLM and lenses have inherent characteristics of parallel processing.
2. Optical encryption methods possess multiple-dimensional and multiple-parameter capabilities. The optical parameters for security keys can be wavelength, polarization and phase.
3. For optical encryption, researchers require multidisciplinary knowledge regarding optical signal processing, image processing, optical theories, and computer technologies as well.

*Applications of optical signal processing*

Fractional transforms and more precisely, fractional Fourier transform have gained keen interest from researchers in the area of optical signal processing. Thus, it's also commonly referred to as "Fourier Optics" or "Information optics". Fractional transforms have a widespread application in signal processing and image processing, in the area of time-variant signal filtering, phase retrieval, image restoration, pattern recognition, tomography, image compression, encryption and watermarking. This article focuses on the image encryption application of various fractional integral transforms.

## 2.2 DRPE Model for image encryption

Double random phase encoding (DRPE) based image encryption has its roots in the work of Refregier and Javidi [47] where two random phase functions in fractional Fourier domains are used to encrypt input plain image into stationary white noise. Hennelly and Sheridan [48] have shown image encryption as random shifting in the fractional Fourier domain. Unnikrishnan [49] has generalized the DRPE scheme in the fractional Fourier domain. The DRPE architecture is most exhaustively used and explored in various optical processing-based applications. The research community has been continuously exploring the possibilities to improve the security of DRPE [50] [51] [52] [53] [54] and has also successfully extended the DRPE scheme to other linear canonical transforms (LCTs) domains. Fig.2 shows the schematic architecture of DRPE based image encryption scheme. As shown in Fig.2, there are two random phase masks (RPM) also known as phase-only masks (POM). One of the POM is placed at the input plane and another is placed at the Fourier plane. The $POM_1$ at the input plane makes the input signal/image white noise-like but nonstationary and $POM_2$ at the Fourier plane is also a white noise but is stationary. Let $POM_1$ at the input plane be $\exp(j\phi(x,y))$ and $POM_2$ at Fourier plane as $\exp(j\varphi(\mu,\nu))$, both being randomly distributed in the range $[0,2\pi]$. Therefore, wavefront after $POM_1$ is given by

$$F(\mu, v) = FT\{I(x, y) \exp(j\phi(x, y))\} \tag{5}$$

where $I(x, y)$ is input image in the spatial domain, $FT$ denotes a Fourier transform operation. The wavefront, $F(\mu, \vartheta)$ gets modified by POM$_2$ in the Fourier domain and an inverse Fourier ($IFT$) is performed over it. This gives a complex domain wavefront as

$$C(\xi, \eta) = IFT\{F(\mu, v) \exp[j\varphi(\mu, v)]\} \tag{6}$$

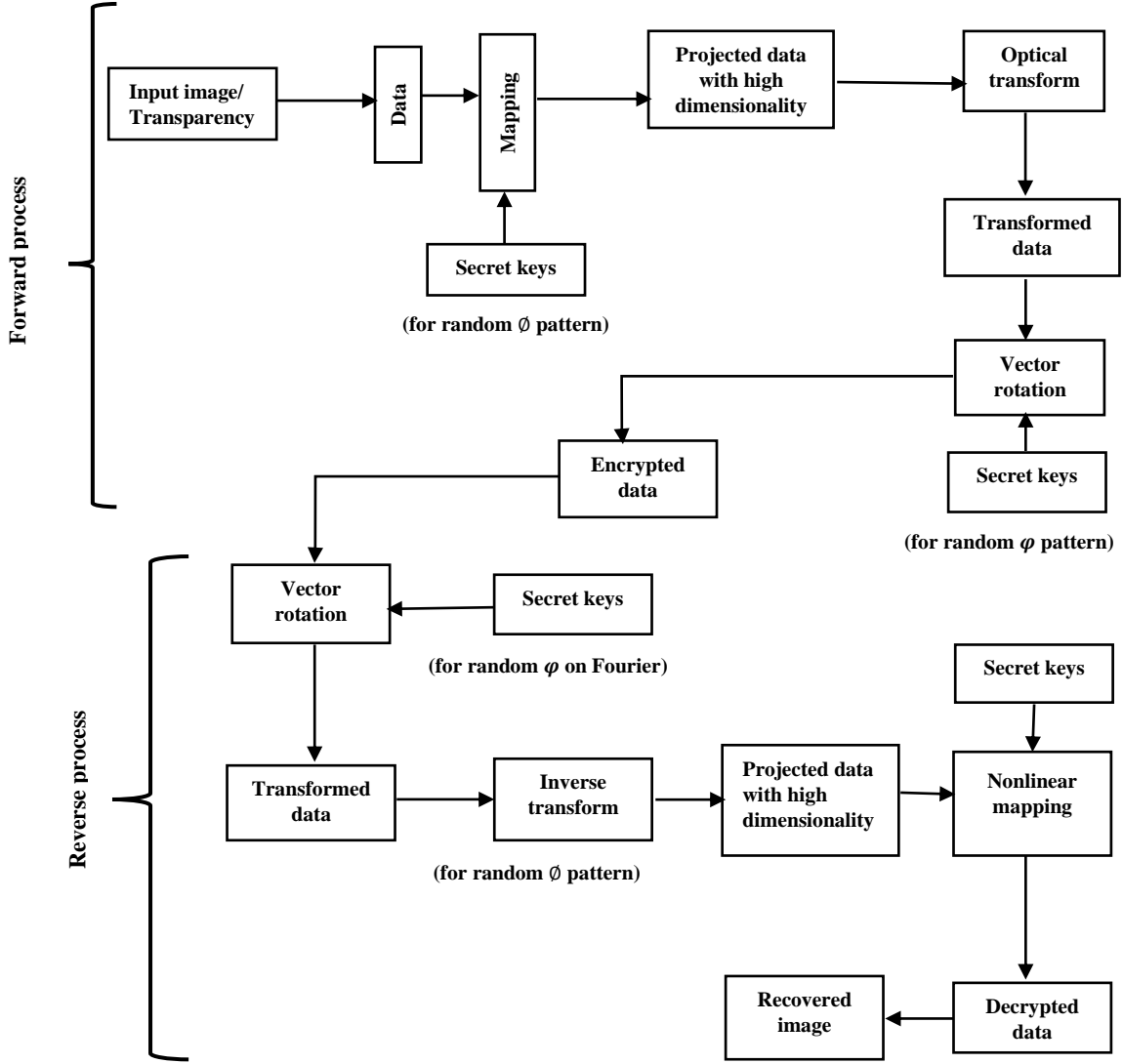

**Fig. 2.** Architectural model for DRPE-based encryption scheme

The complex-valued coefficients are recorded on a CCD (charged coupled device) in optical processing while the terms can be electronically recorded in a computer. During the decryption/reverse process, the complex domain wavefront is first transformed to POM$_2$ as

$$\hat{F}(\mu, v) = \{FT[\hat{C}(\xi, \eta)] \{\exp(j\varphi(\mu, v))\}^* \tag{7}$$

where $\{.\}^*$ represents a conjugate operation. *IFT* of Fourier wavefront is obtained with POM$_1$ conjugate as

$$\hat{I}(x, y) = \{IFT[\hat{F}(\mu, v)]\} \{\exp(j\phi(x, y))\}^* \tag{8}$$

Thus, $\hat{I}(x, y)$ is the decoded wavefront in the spatial domain.

DRPE schemes are broadly classified as 1) Amplitude-only DRPE where decoding is done without using POM$_1$. 2) Fully phase DRPE where the input image is fully converted into a full phase map. This phase-only map is used to encode images with the DRPE procedure. The only difference is that the input image is first normalized and converted into a phase image as $\exp[jI(x, y)]$ before encoding. Details of each classification is beyond the scope of this review work. However, it is likely to mention that each POM at the input as well as Fourier domain can be used as secret keys. This enlarges the key space thereby enhancing security.

### 2.2.1 Previous review articles and contributed evaluations

There are many review articles available in the literature [55] [56] [57] that provide the evolution of classical DRPE-based architecture. Some of the significant contributions in reviewing fractional transforms are listed in Table 2. The contribution of these reviews is summarized on various aspects and evaluations included in them. Each review article is categorized according to the evaluation of various schemes in the work. Whereas some of these are based on just conceptual and theoretical aspects, while others provide an evaluation of quantitative, qualitative, comparative, applications etc. We have nomenclated these evaluations from E01 to E09 based on the criteria mentioned at bottom of Table 2.

**Table 2** Recent review articles on fractional transforms-based image encryption schemes.

| Author[Ref] | Year | Description | Evaluations done |
|---|---|---|---|
| Moreno [55] | 2010 | On the usage of optical signal processing and its conceptual and theoretical details | E01, E08 |
| Sejdić [58] | 2011 | On FrFT digital realizations and related application areas | E01, E05, E06, E09 |
| Saxena [59] | 2013 | On FrFT and its properties, versions in the discrete domain and some application areas | E01, E05, E09 |
| Chen [56] | 2014 | On the advances in optical security, various optical signal processing schemes illustrated | E01, E02, E06, E07, E08, E09 |
| Yang [60] | 2016 | On fractional calculus and MATLAB functions defined for same, various application areas reviewed | E01, E02, E05 |
| Javidi [57] | 2016 | On recent advances and challenges of optical security using free space optics, cryptanalysis and road map to the development of secure theory in optics. | E01, E02, E05, E06, E08, E09 |
| Guo [61] | 2016 | On the vulnerability of LCT-DRPE based encryption to COA with numerical implementation | E01, E02, E03, E07, E08 |
| Situ [62] | 2017 | A review on phase problems in optical imaging | E01, E05, E07, E08, E09 |
| Guo [45] | 2017 | On recent development in iterative phase retrieval and application in information security | E01, E02, E05, E07, E08, E09 |
| Kaurl [63] | 2018 | On the latest developments in the meta-heuristic methods of image encryption | E01, E02, E03, E04, E06, E07, E09 |
| Jinming [64] | 2018 | On research progress in theory and applications of fractional Fourier transform | E01, E02, E05, E06, E07 |
| Gadhrili [65] | 2019 | On different algorithms for color image encryption | E02, E03, E04 |
| Jindal [66] | 2019 | On the applications of fractional transforms in image processing | E04, E07 |
| Gómez-Echavarría [67] | 2020 | On the applications of fractional Fourier transform in biomedical signal processing | E01, E05 |

**E01: Conceptual and Theoretical, E02: Quantitative, E03: Qualitative, E04: Comparative on results, E05: Applications explored, E06: Vulnerabilities, E07: Architecture, E08: DRPE based, E09: Mathematical details**

This will give better clarity to the reader and future researchers regarding various aspects discussed in each review. It is not possible to include all the related work in this paper for the sake of brevity. However, best efforts are put to include the most recent developments in DRPE-based encryption schemes as listed in Table 3. DRPE-based architecture has been extensively used and is considered as an effective method. DRPE methods require a random phase mask as the secret key that needs to be stored at the receiver for decryption. Besides that, a careful alignment of the random phase mask with received encrypted data has to be done. The inherent property of linearity and symmetricity proves to be a bane of encryption applications as the linearity may lead to vulnerability to different types of attacks. Based on these vulnerabilities, some of the recent works on cryptanalysis are summarized in Table 4. Each reference is included with a short description of the work and methodology adopted to cryptanalysis the security scheme.

**Table 3** Recent publications on evolutionary methods adopted in optical transform with DRPE-based architecture (2016-2021)

| Author [Ref], Year | Method | Security | Advantages | Limitations |
|---|---|---|---|---|
| Abd-El-Atty et al. [68], 2021 | Based on the application of DRPE and quantum walks. An alternate quantum walk (AQW) is used to generate random masks as well as for permutation. | Moderate | 1.Higher key space 2.Resistance to digital and quantum computer attacks. | 1.Non uniform histograms 2.Classical attack analysis missing 3.Differential attack analysis not discussed. |
| Zhou et al. [69], 2020 | Image is transformed in DRPE domain. The phase information is | High | 1.Simultaneous compression and encryption. | 1.Higher complexity |

| | | | 2. Faster and efficient.<br>3. Robust to differential attacks | 2.PSNR is lower indicating degraded reconstructed image. |
|---|---|---|---|---|
| quantized for its usage in the authentication. The plaintext is compressed by CS where the measurement matrix is also quantized using a sigmoid function. | | | | |
| Huang et al. [70], 2020 | Low-frequency subbands are extracted by contourlet transform. Scrambled with 2D logistic map. 2DLCT is applied to obtain phase truncation and phase reservation. This is followed by an XOR operation with a logistic map. | High | 1.Multiple image encryption<br>2.Uniform histograms<br>3.optimum entropy and CC of encrypted<br>4.Robust to classical and differential attacks | 1.Performance degrades considerably with data loss and noise attack |
| Wang et al. [71], 2020 | Based on apertured Mellin transform realised by log-polar transform followed by apertured fractional Fourier transform. | High | 1.Key size increased<br>2.Non linearity in transform is able to resist potential attacks | 1.Quality of decrypted images vary with aperture length parameter<br>2.Mellin transform gives a lossy recovery, resulting in significant degradation in recovered image |
| Huang et al. [46], 2019 | Original image is encoded with a modified Gerchberg-Saxton algorithm, which is controlled by hyperchaos system derived from Chen chaotic map. Josephus traversing is used for scrambling the phase function followed by diffusion-confusion by hyperchaos. | High | 1.Uniform histograms<br>2. High sensitivity to keys<br>3. Optimum entropy<br>4.Resistant to all potential attacks | 1. Hyperchaotic map has high complexity in hardware implementation.<br>2. G-S algorithm based on hyperchaos increase encryption/decryption time |
| Huo et al. [72], 2019 | Based on DNA theory with DRPE technique with PWLCM based keys and random phase masks. Initial values of PWLCM are generated by massage digest algo5(MD5). Two rounds of process gives ciphertext. | High | 1.High security to input keys<br>2.key space is large | 1.Axis alignment is required for optical setup<br>2.Lack in differential attack analysis |
| Liansheng et al. [53],2019 | Based on customized data container. Using phase masks that are generated from Hadamard matrix to collect intensities of data containers. After XOR coding, data is scrambled with logistic map | High | 1.Solves issues related to inherent linearity of computation ghost imaging.<br>2. High sensitivity to keys | 1D logistic map has its own limitations |
| Gong et al. [73], 2019 | Based on compressive sensing (CS) and public key RSA algo with optical compressive imaging system to sample input image. Walsh Hadamard transform, followed by scrambling with compound chaos | High | 1.Enlarged key space<br>2.Resistant to CPA<br>3.Entropy is optimum for both global and local values<br>4.Robust to noise and data loss attack | 1. Higher complexity for implementation |
| Chen et al. [74], 2019 | Chaotic Ushiki map is used to generate random phase masks. A single intensity image is encrypted from color image. An equal modulus decomposition used to create asymmetric keys | High | 1.Enhanced security by Ushiki chaotic map<br>2.Enlarged key space<br>3. Immune to CPA and KPA | 1.Lossy recovery<br>2.Entropy not reported<br>3.Differential attack analysis not done |
| Yadav et al. [75], 2018 | Input is first transformed with chaotic Arnold transform. Phase masks are based on devil's vortex Fresnel lens (DVFL) | High | 1.Use of DVFL eliminates axis-alignment issues.<br>2. Parameters of DVFL, orders of FrHT and AT serve as secret key | Robustness to classical and differential attacks not presented |
| Faragallah et al. [52],2018 | Arnold transform is used to scramble RGB of image followed by a Fresnel based Hartley transform from random phase masks generated with a Logistic adjusted sine map | High | 1.Enhanced security due to enlarged key size<br>2. limitations of logistic map are eliminated<br>3.Optimal CC of encrypted | 1. Histograms are not independent of plane image input to some extent<br>2.UACI=0<br>3. Leakage of information due to low entropy values |
| Kumar et al. [76], 2018 | security key generated from a phase retrieval algorithm is used obtain 2D non-separable linear canonical transform of complex image formed by combining two plane images | High | 1.Double image encryption with asymmetric keys<br>2.Robust to data loss attack<br>3.Chosen plain text attack addressed | 1.Phase retrieval has its inherent complexity |
| Jiao et al. [77], 2017 | QR (quick response) code for speckle noise removal in Fresnel based optical transform | High | 1.Speckle noise reduced in optical transformed output | 1.Applicable only to gray scale images |

| Khurana et al. [78],2017 | Phase truncated Fourier and discrete cosine transform (PTFDCT) with random phase as keys. Decryption requires a cube root operation | High | 1.Robust to differential attack 2. Enhanced security 3.Enlarged key space | 1. Entropy is less than optimum 2. Correlation plots show unequal distributions along both dimensions leading to information leakage. |
|---|---|---|---|---|
| Su et al. [79],2017 | Chaotic phase masks for cascaded Fresnel transform holography and constrained optimization for retrieval | Moderate | 1.Reduces retrieval time by using constrained optimization 2. Key sensitivity high due to use of chaotic Henon map | 1. decrypted image is considerably deteriorated 2. performance will degrade under noisy and occlusion attacks |
| Li et al. [80], 2016 | Depth conversion integral imaging and hybrid cellular automata (CA) | High | 1.PSNR of reconstructed images degraded with noise are higher 2.Key space is high (multidimensional) 3.Good resistance to data loss attack | 1. Lossy decryption 2.Differential attack analysis not proved |

**Although certain probable drawbacks/limitations are mentioned corresponding to each scheme, some specific solutions like security enhancement methods can be applied in practice.**

**Table 4** Cryptanalytic approaches in optical/DRPE based encryption schemes (2016-2021)

| Author | Year | Description | Methodology/ strategy |
|---|---|---|---|
| Guo et al [61] | 2016 | Phase retrieval attacks on LCT based DRPE schemes | Hybrid input–output algorithm, error reduction algorithm, and combinations of both type of phase retrieval algorithms are applied for ciphertext-only attacks on Separable LCT DRPE system. |
| Yuan et al [82] | 2016 | Cryptanalysis and its remedy in encryption based on computational ghost imaging | Due to linear relation between input and output of the encryption with Computational ghost imaging is attacked |
| Li et al [83] | 2016 | Vulnerability of impulse attack-free DRPE scheme to chosen plaintext attack | CPA on impulse attack free-DRPE is breached by using a new three-dimensional phase retrieval algorithm. |
| Wang et al [84] | 2016 | Cryptanalysis in phase space | Phase space information vulnerable to chosen plaintext attack (CPA) and known plain text attack (KPA). |
| Liao et al [85] | 2017 | Ciphertext only attack on optical cryptosystem | Based on autocorrelation between plaintext and ciphertext, COA is imposed. |
| Hai et al [86] | 2018 | Cryptanalysis of DRPE scheme with deep learning | Vulnerability to CPA with working mechanism-based learning with neural network. |
| Xiong et al [87] | 2018 | Cryptanalysis of optical cryptosystem with combined phase truncated Fourier transform and nonlinear operations | A phase retrieval attack with normalization and bilateral filter is proposed. |
| Dou et al [88] | 2019 | Known plaintext attack in JTC-DRPE scheme | Application of denoising operations make the cryptosystem linear. Thus, KPA is possible. |
| Xiong et al [89] | 2019 | Cryptanalysis in optical encryption based on vector decomposition of Fourier plane | Cascaded EMD (equal modulus decomposition) based cryptosystem is attacked with CPA and a special attack. |
| Chang et al [90] | 2020 | Ciphertext only attack in optical scanning cryptography (OSC) | A linear system property analysed in the ciphertext expression equation of OSC lead to COA. |
| Jiao et al [91] | 2020 | Known plaintext attack in cryptosystem based on space and polarization encoding | Matrix regression based on training samples is proposed to crack a space based optical encoding and double random polarization encoding with KPA. |
| Zhou et al [92] | 2020 | Vulnerability of encryption scheme based on diffractive imaging to machine learning attacks | An end-to-end machine learning strategy is adopted to establish relationship between ciphertext and plaintext in case of diffractive imaging. |
| He et al [93] | 2020 | Cryptanalysis of optical cryptosystem using untrained neural network | Untrained NN is used to break a phase truncated Fourier transform based optical asymmetric crypto system. Parameters are optimized by plain-ciphertext encryption model of phase truncated Fourier transform. |
| Song et al [94] | 2021 | Cryptanalysis of phase only information as it is vulnerable to chosen plaintext attack. | Deep learning structure is trained using sparse phase information of the encrypted domain image as phase only information is vulnerable to classical attacks. |
| Li et al [83] | 2016 | Vulnerability of impulse attack-free DRPE scheme to chosen plaintext attack | CPA on impulse attack free-DRPE is breached by using a new three-dimensional phase retrieval algorithm. |
| Wang et al [84] | 2016 | Cryptanalysis in phase space | Phase space information vulnerable to chosen plaintext attack (CPA) and known plain text attack (KPA). |
| Liao et al [85] | 2017 | Ciphertext only attack on optical cryptosystem | Based on autocorrelation between plaintext and ciphertext, COA is imposed. |
| Hai et al [86] | 2018 | Cryptanalysis of DRPE scheme with deep learning | Vulnerability to CPA with working mechanism-based learning with neural network. |

## 2.3 Mathematical Modelling of optical transforms with FRFT and its variants

**Table 5** Various methods for discretization of Linear Canonical transforms

| Type | References | Pros | Cons |
|---|---|---|---|
| Sampling type DFrFT | [95] | A direct and simplest of all methods | discrete version is derived at the cost of losing many important properties like unitary, reversibility and additivity. Therefore, has limited applications |
| Improved Sampling type DFrFT | [9] | It works like a continuous FrFT and is a fast algo | Doesn't have orthogonal and additive property. Also, it requires to put some constraints on input signal. |
| Eigen vector decomposition based DFrFT | [10] [12] [16] [17] | Based on eigen values and eigen vector of DFT matrix and then evaluating their fractional power. Retains orthogonality, reversibility, and additivity. Further improved by orthogonal projection in [12] | This type of DFrFT lack fast computation, and the eigen vectors cannot be written in closed form. |
| Linear combination type DFrFT | [13] [19] [20] | Eigen vectors are derived by linear combination of identity operation, DFT, time inverse operation and IDFT. Satisfies properties of reversibility, additivity and orthogonality. | The outcome of transform does not match with continuous transform. It works very much similar to Fourier transform and lose characteristics of fractionalization of powers. |
| Chirp type DFrFT | [96] | DFrFT is derived as multiplication of DFT and periodic chirp signals. Satisfies additivity, reversibility property along with Wigner distribution's rotation property. | There are constraints on the selection of rotation angles and also $N$ (sample length) should not be a prime number. This makes it complicated |
| Closed form DFrFT | [15] | Derived 2 types of DFrFT and Discrete Affine transform (DAFT). Performance is similar to continuous FrFT for Type I and can be calculated using FFT. Type II is improved form of Type I and is applicable to signal processing. Has lowest complexity. | Scaling property exists for only Type I and not for Type II. |

    Linear canonical transforms (LCT), time-frequency transforms and fractional Fourier transform (FrFT) are closely related. Since the application of FrFT to signal processing is proposed [4] [5] [8], there has been tremendous development in the application of FrFT and its variants to image encryption. As fractional transform orders serve as the secret key, the digital implementation is particularly suitable for encryption applications [48]. Since this work is mainly focussed on the application of fractional transform in image encryption only, we won't elaborate the mathematical eloquence behind the fractional transforms here. This section specifically emphasises the discrete realizations (DFrFT) and their application to image encryption. There are various methods proposed in the literature for the discretization of fractional transforms, some of them are classified as shown in Table 5 with pros and cons of each type. It is worth noting that Table 5 includes only a fractional version of Fourier transform. This is due to the fact that the fractionalization of Linear canonical transforms started with Fourier transform itself and later was extended to other transform domains. The methods of discretization mentioned below are therefore conceptually applicable to variants of Fourier transforms as well viz. Gyrator transform [97, 50], Mellin transform [98, 99, 100], Hillbert tranform [101], Hartley transform [17] [20], Hadamard transform [19] etc.

Fig.3 shows the basic architecture for fractional transform-based image encryption that is digitally implemented without a random phase mask in either domain (without DRPE). As depicted in Fig.1, this method requires the knowledge of fractional transform orders that are used along both dimensions within a range [0,1]. The decryption is exactly similar to the forward process and requires the same fractional orders but with negative values to decrypt the image correctly. The encryption is thus a symmetric scheme and a slight change in the key value will result in incorrect decryption.

    The major limitation of such a scheme is shorter key space which makes it vulnerable to brute force attacks. The input image is pre-processed for enhanced security and enlarging a key space. The pre-processing can be a scrambling operation that only shuffles the pixel positions to make the image, unintelligible. In some cases, this pre-processing can be a nonlinear operation that can be a substitution of pixel intensity values. There are various schemes that employ either scrambling [102] [103] [104] [105] , substitution [106] or both [51] [107] [108] to enhance the security. The following section includes all major schemes that are proposed to improve the performance of fractional transform-based image encryption. We have categorized them in accordance with the strategical amalgamation of scheme with fractional transform domain. The schemes proposed in the literature are nomenclated in eight major categories (T01 -T08). Each amalgamated scheme is reviewed separately. This portion

of review article is elaborated as our emphasis is on the digital implementation of fractional integral transforms for image encryption.

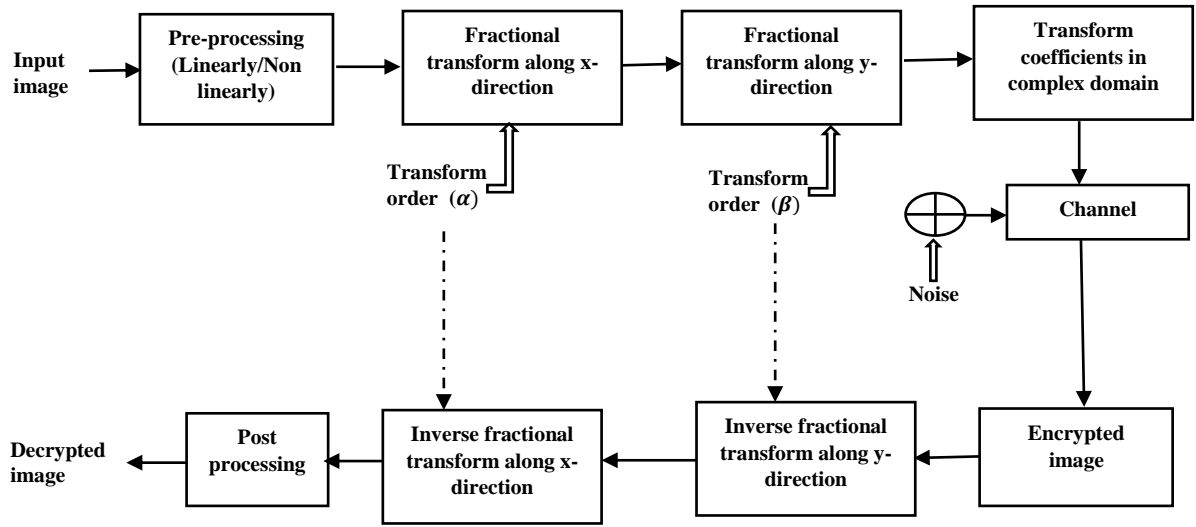

**Fig.3.** Schematic architecture for Fractional transform-based image encryption in digital domain

*2.3.1    Reality preserving with optical transform domain (T01)*

The optical transform results in complex coefficients output corresponding to a real domain input image. Although it is easy to process these complex coefficients with a holography method but in a digital domain, it requires two images to be processed in the encrypted domain, one for real terms and other for imaginary terms. Therefore, storage and transmission increase complexity and overheads in digital channels. To overcome this limitation, Venturi and Duhamel [109] proposed a mathematical solution based on the properties of the complex transform output. Reality preserving refers to real domain output for a real domain input signal. The algorithm still has computational complexity, $O(N^2)$ for matrix order of *N*. Reality preserving transforms that are formulated with this algorithm have most of the required properties of fractional transforms along with a monotonously decreasing decorrelation power. Such transforms are beneficial where orthogonal reality preserving transform is required with their decorrelation power controlled by some parameter such as, in joint source and channel coding. Initially, the algorithm was proposed in fractional sine and cosine transforms. It is further extended to other transforms with the basic properties of the transforms retained well. Recently Zhao et. al [99, 110] used it to obtain fractional Mellin transform for triple image encryption. Reality preserving is also used in discrete fractional Cosine transform [111, 105], fractional Angular transform [112, 113], fractional Hartley transform [114, 115, 116, 117], besides  fractional Fourier transform [103, 104, 118].

*2.3.2    Application of chaos theory in optical transforms-based image encryption (T02)*

  Chaos theory refers to the study of unpredictable behaviour in systems governed by deterministic laws. Chaotic properties are closely related to cryptography [119] owing to their sensitivity to initial conditions, randomness and ergodicity. Due to such intrinsic characteristics, chaotic maps have been extensively used in data encryption. Chaotic maps are used as pseudorandom generators [120], for substitution, and permutation of image pixels. Various schemes for encryption based on permutation only [121, 122],  or substitution only  [123]   or a combination of both [120], [107] with the usage of either one-dimensional basic maps like logistic [124], sine, the tent [125], 2D Chirikov standard map [120], or higher dimensional compound chaos or higher dimensional hyperchaotic maps [126] [127] [128], depending on the application and level of security.
  Chaotic maps have been extensively used in amalgamation with optical transforms-based image encryption for enhancing security. Fractional transform-based image encryption schemes have only transform orders as the secret key. However, this key space is not large enough and is therefore vulnerable to cryptanalysis. To enhance security, chaotic maps are used that also enlarge the key space. There are various schemes proposed in the literature that have used permutation with chaotic maps along with an optical transform [50] [129] [51] [52]    [104] [103] [130] [131]. The order in which these two schemes are amalgamated may vary. Permutation in the spatial domain followed by transform or transform followed by permutation in the transform domain. Some of the schemes follow substitution-permutation and transform collectively [107] [132] [133] to further enhance security. We have reviewed some of the most recently proposed schemes that use chaos-based permutation/substitution with optical transforms.

Wu et al [108] proposed a color image encryption scheme in random fractional discrete cosine transform (RFrDCT) along with scrambling and diffusion paradigm (DSD). A logistic map is used to generate a randomized vector of fractional order. This enlarges key space and increases sensitivity.

A multiple parameter fractional Hartley transform (FrHT) is proposed by Kang et al. [114] with its reality preserved for a color image encryption. The chaos is embedded into the algorithm at each step. The original color image with individual color components is first combined into a single image. This single image is divided into different sub-blocks. The blocks are then shuffled based on a pseudo-random sequence generated from non-adjacent coupled map lattices (NCML) based on logistic maps. The initial parameters of NCML are generated from yet another chaotic map (Arnold Cat map). The initial parameters of chaotic maps at this stage serves as secret keys. Next stage of encryption is based on a pixel scrambling operator which is based on a 2D Chirikov standard chaotic map (CSM). Using CSM, a series of 2D and 3D angle matrices are generated that are used to convert images in RGB space to newer space. The final stage is to obtain an MPFrHT in real domain (RPMPFrHT) and to divide the image into three to get concatenated encrypted image as ciphertext.

A new fractional transform coined as the non-separable fractional Fourier transform is proposed by Ran et al. [134]. Random phase masks are generated by Arnold transform. The advantage of this type of transform is that it is able to tangle information along and across two dimensions together. It is closely related to the Gyrator transform. Also, the proposed scheme is resistant to decryption with multiple keys, unlike ordinary fractional Fourier transform.

Wu et al. [135] proposed a reality preserving fractional Discrete cosine transform (RFrDCT) for image encryption. The RFrDCT domain image is subjected to confusion-diffusion paradigm. The confusion is obtained using a game of life algorithm (GoL) and diffusion in the next stage is based on an *XOR* operation with another chaotic map. The initial parameters of chaos serve as secret keys of encryption. Enhanced performance is claimed with the adopted strategy. A perturbation factor is applied for resistance against differential attacks.

An encryption scheme with S-box generation is proposed in [131] which is unique in the way these S-boxes are generated. Chaotic Chebyshev map and linear fractional transform are used for the construction of S-box. Partial image encryption is achieved by a permutation-substitution-diffusion (PSD) network and multiple chaotic maps in the linear wavelet transform (LWT) domain. Using dynamic keys for controlling encryption aids in security against differential attacks. Partial encryption of only sensitive portions not only reduces computation complexity but is also faster and more efficient.

Jamal et al. [136] proposed yet another scheme that uses a combination of linear fractional transform and chaotic systems to generate substitution boxes for image encryption. The chaotic maps used in the scheme are generated from a combination of seed maps to enhance the security and chaotic range. The investigation for complexity thus obtained with the proposed scheme is based on various algebraic and statistical tests. The investigation gives testimony of improved perplexity and confusion in the encrypted domain.

A novel Fresnel-based Hartley transform is proposed in [52] for an optical-double color image encryption scheme. The color image is first separated into individual channels and are scrambled separately with the Arnold transform (AT) in spatial domain. Each scrambled image is then multiplied with a 2D chaotic Sine- adjusted logistic map (LASM) and then a Hartley transform is applied to each channel. This procedure is repeated once again with another set of AT based scrambling (now in Hartley domain) and then each channel is multiplied with another set of 2D-LASM. The final step is obtaining inverse Hartley transform which gives an outcome across each channel in Fresnel domain. The color channels in Fresnel domain are concatenated to obtain a single image which is the final ciphered image.

A fractional angular transform (FrAT) is used in [137] where plain image is substituted with a chaotic logistic map prior to transform. The transform orders along with initial value of logistic map serve as secret keys of encryption. The scheme performs marginally as there are certain limitations due to similarity in histograms of plain and encrypted domain and correlation coefficients in encrypted domain are considerably higher. Moreover, the scheme is not evaluated for entropy measure and differential attack analysis.

### 2.3.3   *Compressive sensing (T03)*

Compressive sensing (CS) also refers to sparse signal sampling, was introduced by work of Donoho, Candes [138, 139]. CS is able to achieve compression and signal sampling simultaneously [73] [140] [141] . For a signal of bandwidth, $BW = \Omega$, the sampling frequency ($f_s$) required to represent the signal is much smaller than Nyquist frequency ($f_s \ll \Omega$). Let $R^N$ be the set of N-tuples of real numbers. If $x \in R^N$ is input 1D signal sampled using CS, then $x$ can be sparsely represented using an appropriate basis function $\Psi = [\psi_1, \psi_2 \ldots \psi_N]$. Thus $x = \Psi_s = \sum_{i=1}^{N} s_i \psi_i$. Let $y_{M \times N}$ be the measured matrix with $M \ll N$. Then $y = \emptyset x = \emptyset \Psi_s = As$ where $y \in R^N$. Thus if measurement matrix, $A$ that is used to measure sparse signal, $s$ is given, then the construction of signal requires solving an under determined linear system and the sparse signal can be obtained by solving a combinatorial optimization problem given by : $min\|s\|_0 : y = \emptyset\, \Psi_s = A\, s$.

A collective compression-encryption scheme is proposed in [100] with 2D compressive sensing and fractional Mellin transform. The original image is first measured using a measurement matrix in both dimensions to reduce data volume with 2D CS. The measurement matrix is constructed using partial Hadamard matrices. Chaos is used to control the measurement matrix with its initial conditions. The non-linear Mellin transform is used to overcome the security issue related to linear transform.

Zhao et al. [142] proposed a double-image encryption scheme which is claimed to be faster and more efficient. The scheme utilizes DWT as the basis for the measurement matrix. Both images are first transformed into discrete wavelet transform (DWT) basis and are compressed with the measurement matrix derived from 2D Sine-Logistic modulation map (2D-SLMM). The images are then combined and Arnold transformation is applied for scrambling the coefficients. Two circular random matrices are generated using 2D-SLMM with different seed values. These random matrices are used to obtain discrete fractional Random transform (DFrRT). The encrypted image is thus in DFrRT domain.

In another CS based scheme proposed by Zhang et al. [143], Kronecker product (KP) is combined with the chaotic map for the generation of measurement matrix and random phase masks. Low dimensionality seed maps are extended to high dimensional by KP. These high-dimensional maps are used for the measurement matrix. The scheme is able to provide an efficient and fast approach to color image encryption.

A comparatively simpler scheme is proposed in [144] where image compression-encryption uses a combination of 2D CS and discrete fractional random transform (DFrRT). The basis function for the measurement matrix is a DCT (discrete cosine transform). The measurement matrix is constructed with a chaotic logistic map to control row vectors of the Hadamard matrix. The compressed image is then encrypted by DFrRT. Reconstruction of CS requires Newton's smoothed $l_0$ norm ($NSL_0$) algorithm.

An asymmetric cryptosystem for color images based on CS and equal modulus decomposition (EMD) is proposed by Chen et al. [145]. In this scheme, the color image is initially combined to a single image. With the application of DWT, this image is converted into low-frequency and high-frequency images. The high-frequency image is compressed by a measurement matrix generated from logistic map. The compressed image is segmented into two matrices. One of the matrices is used as a private key (a random matrix related to the plain image) for DFrRT and another matrix is combined with the low-frequency image to form a complex function. This complex function is transformed into DFrRT with the private key (random matrix) that is plain image dependent. This enables the cryptosystem to resist known and chosen plaintext attacks. The output of DFrRT is decomposed into 2 masks by using EMD where one mask is a cipher image and another is a private key. The inverse CS in the decryption process is based on the basis pursuit algorithm (BP).

Yi et al. [146] proposed to use multiple measurement matrices instead of a single measurement matrix that is used to sample all blocks of an image. This strategy enables to overcome the issue of chosen plaintext attacks. The mother measurement matrix is derived from a single chaotic map and other measurement matrices are generated by exchanging rows using a random row exchanging method. However, another chaotic map is required to control the row exchanging operation. The compressed image is then transformed with fractional Fourier transform (FrFT). The transform is followed by two consecutive pixel scrambling operations to guarantee nonlinearity and to increase key sensitivity in the proposed scheme. Ye et al. [147] proposed a compressed sensed color image encryption scheme based on quaternion discrete multi-fractional random transform with the hash function SHA-512. The parameters of chaos are updated by randomly selected hash values. The use of multifunctional transform not only increases the key space but also improves the key sensitivity.

*2.3.4 On the basis of fixed/multiparameter (T04)*

Fractional transforms can decorrelate the spatial domain pixels based on the fractional value of the transform orders. The fractional transforms are also looked upon as Wigner distribution where each fractional order corresponds to an angle of rotation in the optical domain [4]. With a fixed value of transform orders, the key space is limited and the cryptosystem is vulnerable to brute force attack. To overcome this limitation, various researchers proposed to use multiple parameter-based fractional transforms [148, 149, 150, 151, 152, 132, 153] with their own definitions and postulates. Mathematically, a fractional Fourier transform has multiplicity which is due to different choices of both eigenfunction and eigen value classes [148]. Thus, the multiplicity is intrinsic in a fractional operator. Lang [118] proposed a multiparameter fractional Fourier transform where the periodicity of $M$ is utilised. The transform order vector, $n$ can be $M$- dimensional integer vector. This provides an extra degree of freedom as the periodicity parameter; $M$ serves as a secret key along with the vector parameters.

Sui et al. [154] proposed a multiparameter discrete fractional angular transform (MPFAT) for image encryption that uses fractional order and periodicity parameters to provide multiple parameters in the transform. Similar to a discrete fractional Angular transform (DFAT), MPDFAT also satisfies properties like linearity, multiplicity, index additivity etc. Zhong et al. [155] proposed a discrete multiple parameter fractional Fourier transform (DMPFrFT) for image encryption by using the periodicity parameter for extending to multiple parameters.

Azoug et al. [51] proposed yet another opto-digital image encryption with a multiple parameter discrete fractional Fourier transform after a non-linear pre-processing of the image in spatial domain with a chaotic map. The multiparameter scheme is extended based on the work of Pei et al. [156] which extend the DFrFT to have multiple order parameters equal to the number of input data points. If all the parameters are made equal in an MPDFrFT then it reduces to a single parameter DFrFT.

A general theoretical framework of MPDFrFT is presented in [132]. The work proposed two different frameworks as Type I and Type II MPDFrFT that include existing multiparameter transforms as their special cases. Further, an in-detail analysis of the properties of such transforms is discussed and higher dimensional operators are also defined. Some new types of transforms such as MPDFrCT, MPDFrST, MPDFrHT (Cosine, Sine, Hartley) are constructed under the proposed framework along with their applications such as feature extraction and 2D image encryption.

A quaternion algebra is used with multiple parameter fractional Fourier transform (MPFrQFT) by Chen et al. [106] for generalising MPFrFT. Both forward and reverse MPFrQFT transform are defined and a color image encryption based on the proposed transform is evaluated for its performance as compared to other encryption algorithms. The proposed scheme has larger key space and is more sensitive to transform orders.

Ren et al. [157] proposed a multiple image encryption scheme based on discrete multiple parameter fractional Fourier transform (DMPFrFT) for which original images are filtered in DCT domain and multiplexed into a single image. The multiple parameters are again generated using a periodicity parameter which serves as one of the keys. Other keys are the parameters for scrambling the multiplexed image (random matrix), and transform orders of DMPFrFT.

A multiparameter discrete fractional Hartley transforms for image encryption is proposed by Kang and Tao [114]. The multiple parameters are generated by extending the fractional order to N-dimensional vector and the FRHT kernel is represented as a linear summation with weighting coefficients.

### 2.3.5 DNA Sequence (T05)

DNA coding method is inferred from the Deoxyribonucleic acid and is a branch of computing based on DNA, biochemistry and molecular biology hardware. DNA sequences appear in the form of double helices in living cells. A DNA code is simply a code of alphabetic set $Q = \{A, T, C, G\}$. These alphabets refer to 4 nucleic acid bases: $A$ (adenine), $C$ (cytosine), $G$ (guanine), $T$ (thymine): $A$ and $T$, $G$ and $C$ are complimentary. The complimentary rules are referred to as Watson-Crick compliment [158]. Thus, pairing can be described as: $\bar{A} = T, \bar{T} = A$, $\bar{C} = G$, $\bar{G} = C$ and if a binary code is given to each as 00,11,01,10 with (00,11) and (01,10) as complimentary. With vector algebraic operations based on DNA computing [159] [160], pixel permutation and substitution can be performed if the image pixels are represented in the form of binary sequences.

Recently Farah et. al [102] proposed to use FRFT along with chaos and DNA for image encryption. Initially, a random phase matrix is generated using a chaotic Lorenz map. The plain image is converted to a binary matrix and encoded according to chosen DNA encoding rule. Also, the random phase matrix is encoded to DNA sequence with the same rule. The coded plain image is *XOR*ed with that of the encoded random phase matrix. Using the random phase masks generated from the 3D chaotic map (Lorenz map), iterative FrFT is performed and the resultant image is XORed with the third chaotic sequence to obtain the final ciphered image.

An optical image encryption set -up based on DNA coding is proposed by Huo et al. [72] where a piecewise linear chaotic map (PWLCM) is used to generate a key matrix as well as a random phase matrix. A message digest hash algorithm (MD5) is used to generate initial values of PWLCM. An MD5 hash of plaintext consists of 128 bits. *XOR* operation for DNA is used. Initially, the plain image and key matrix are converted to binary sequences with DNA coding rules that are different for different rows in the image. The DNA encoded plain image is *XOR*ed with a key matrix and a forward Fresnel domain DRPE is applied to obtain the final ciphered image.

### 2.3.6 Cellular Automata (T06)

Cellular Automata (CA) also called cellular spaces, tessellation automata/structures, cellular structures or iteration arrays find application in various fields like physics, microstructure modelling etc. CA consists of regular rigid cells that are generated in accordance with a fixed rule which is nothing but a mathematical function. CA is used in cryptography due to the possibility of pseudo-random number generation with such rule (Rule 30) which is a class III rule displaying aperiodic chaotic behaviour [161] [162]. Li et. al [163] proposed a 3D image encryption by using computer-generated integral imaging (CIIR) and cellular automata transform. An elemental image array (EIA) recorded by light rays coming from 3D image is mapped according to a ray- tracing theory. An encrypted image is then generated from 2D EIA by using cellular automata transform. It is claimed that CA-based encryption is error-free and being an orthogonal transformation, it offers simplicity. The performance of the scheme is measured in terms of BCR (bit correct ratio) and PSNR for reconstructed and is compared to some similar proposed schemes. This scheme of combining optical transforms to that of CA is unique in its methodology. Recently, there is no further exploration of the proposed idea.

### 2.3.7 Double image (T07.1)/Multiple Image (T07.2)

Double image encryption schemes are aimed to provide more efficiency in terms of resources. A double image is simultaneously encrypted and decrypted. Such schemes also provide higher speed and better sensitivity besides less storage space requirement. Therefore double image encryption schemes have drawn attention of various researchers [164] [130] [165] [154] [104] [142].

Recently, Yuan et al. [166] proposed an image authentication with double image encryption based on non-separable fractional Fourier transform (NFrFT). The two images are combined to form a complex image matrix and is transformed with NFrFT. The output of the transform is also a complex matrix. The transform orders and coefficient parameters serve as secret keys. Novelty of the proposed work is in the selection of a partial phase that is reserved for decryption. A nonlinear correlation algorithm is to authenticate the two recovered images. The cross-correlation of two compared images is referred to as non-linear correlation (NC) whose strength is specified by a parameter, $k \in [0,1]$. An appropriate value of $k$ is selected to authenticate the images. Peak to correlation energy (PCE) is a ratio of maximum peak intensity value and total energy of the non-linear correlation plane. Thus, PCE is measured to determine $k$ and hence authenticity.

A double image encryption scheme based on interference and logistic map is proposed in [167] to overcome the silhouette problem. The two input images are initially joined to make an enlarged image. This joined image is subjected to scrambling based on chaotic sequence generated from a logistic map. Then the scrambled image is again separated into two. One of the images is directly used to generate two-phase keys/masks based on optical interference. Another scrambled image is encrypted with DRPE method using first phase mask (key). This is followed by multiplying the complex outcome with another phase mask for transformation to the ciphertext. The author suggests to use input parameters of the logistic map, wavelength and axial mask as secret encryption keys to further enhance the security.

Singh et al. [54] proposed a full phase encryption scheme for its better security compared to amplitude image. The scheme uses two spatial domain input images and converts each of them to a phase image. The phase images are then multiplied with random phase masks (RPM) and transformed in the Gyrator domain with rotation angle, $\alpha$. The gyrator domain images are then added and subtracted to get two intermediate images. The intermediate images are then bonded with structured phase masks based on the Devils vortex lens (DVFL) specified with certain parameters. This is followed by another Gyrator transform with a different rotation angle, $\beta$ to obtain two encrypted images. Decryption is exactly the inverse of the encryption process.

Similar to double image encryption schemes there is another category where multiple images are simultaneously encrypted to reduce the key space as compared to the data to be encrypted (images ) but at the cost of increased complexity [168] [129]. Recently Sui et al. [169] proposed a double image encryption where two images are initially combined into a single image along the column of the first image followed by the second image. This combined image is scrambled with a 2D sine logistic modulation map. Next, the scrambled image is divided into two components to constitute a complex image. One of the components is the phase part and another part is the amplitude of the complex image. The complex image is shared using Shamir's three-pass protocol where the encryption function is a multiparameter fractional angular transform which is preferred for its commutative property.

Sui et al. [170] proposed multiple image encryption with asymmetric keys in the fractional Fourier transform domain. Initially, a sequence of chaotic pairs is generated by using symmetrically coupled logistic maps. This chaotic sequence is used to scramble the spatial domain images. Phase only function (POF) of image is retrieved by using an iterative process of FrFT domain. In the next stage, all the POFs are modulated into an interim which is transformed to real-value ciphertext by FrFT and chaotic diffusion. The three random phase functions are used as keys to retrieve POFs of plain images and three decryption keys are generated in the encryption process.

A multiple image encryption scheme is proposed [171] by combining a non-linear fractional Mellin transform with a fractional Cosine transform. Fractional Mellin transform is used for its robustness to classical attacks. The original images are simultaneously transformed into a discrete Cosine transform domain (DCT) and then re-encrypted with amplitude and phase encoding. The transformed images have changed centre-coordinates due to fractional Mellin transform since FrMT is a log-polar transform of the image followed by a fractional Fourier transform of log-polar image. The fractional orders of FrFT, phases $\psi_j, \theta_j$ are the secret keys.

Recently, Guleria et al. [172] proposed to encrypt three RGB images simultaneously by using RSA cryptosystem followed by a discrete reality preserving fractional cosine transform and the final stage of scrambling with Arnold transform. To accomplish multiple image encryption, 3 RGB images are combined into a single image by using a single color component of each image as R,G,B components. All three indexed images are individually ciphered with the proposed algorithm and then combined as a single ciphered image. The security of the scheme depends not only on the input parameters of RSA, Arnold transform and orders of transform but also on their sequence of arrangement. Decryption is exactly the inverse of the encryption scheme.

*2.3.8 Watermarking in the encrypted domain (T08)*

Recently many researchers have proposed to use of optical transform for watermarking applications [129] [173] [174] [175] [176]. Watermarking an image is a data hiding method for copyright protection and copy prevention. Depending on the application, a watermark can be a visible pattern or can be hidden in the host image. For copyright, its generally a visible pattern and for resolving an authorship problem, the watermark is secretly embedded into image which can be recovered by an authorized user only. In the latter case, the watermark is usually a binary logo that is encrypted into a noise-like pattern and then embedded in the image for enhanced security. Many researchers have followed this approach in the watermarking algorithm. Some of the recent watermarking schemes with an encryption algorithm using fractional transforms are reviewed in this section.

Singh et al. [177] proposed to embed an encrypted watermark in fractional Mellin transform (FrMT) into the host image. The two deterministic phase masks (DPM) are generated to be used in the input and frequency plane. The watermark image is first converted into a log-polar image. After multiplying the log-polar image with the first DPM, it is transformed to a fractional Fourier transform domain. This is FrMT transformation. In the next step, again the second DPM is multiplied by the complex outcome and inverse FrFT is obtained. For embedding the outcome is attenuated by a factor and then added to the host image. SVD decomposition is applied in the last stage to make the watermarked image unrecognizable and is transmitted as individual S, V, D matrices.

A quaternion algebra is used to define a quaternion discrete fractional random transform (QDFRNT) which generalizes DFRNT for its application in watermarking [178]. The host image is divided into blocks and QDFRNT is applied to each block. The scrambled watermark image is used to modify the mid-frequency coefficients of the QDFRNT host image. The transform orders and parameters of the scrambling scheme in the watermark image are used as secret keys of encryption.

Liu et al. [179] proposed a novel transform, known as fractional Krawchouk transform (FrKT) to generalize the Krawchouk transform. Derivation of FrKT is based on eigenvalue decomposition and eigen vectors. For validating the imperceptibility of the proposed transform, a watermarking application is illustrated in the work. A better robustness and imperceptibility with proposed transform have been claimed in the work.

### 3. Performance metrics for image encryption

Image data has high redundancy and large volumes as compared to text or binary data. It may also have some real time operations or may also be incorporated with compressed data of a certain format. Thus, an image encryption scheme needs to satisfy certain requirements. Some of the commonly used performance requirements are discussed in this section. The categorization of such performance analysis is shown in Fig. 4. Performance analysis of encryption requires a comprehensive investigation of perceptual security and cryptographic security. Perceptual analysis requires that the outcome of an algorithm is unintelligible to human perception whereas cryptographic analysis refers to the ability of the algorithm to resist cryptanalysis that includes all possible attacks in terms of the secret key, data statistics etc.

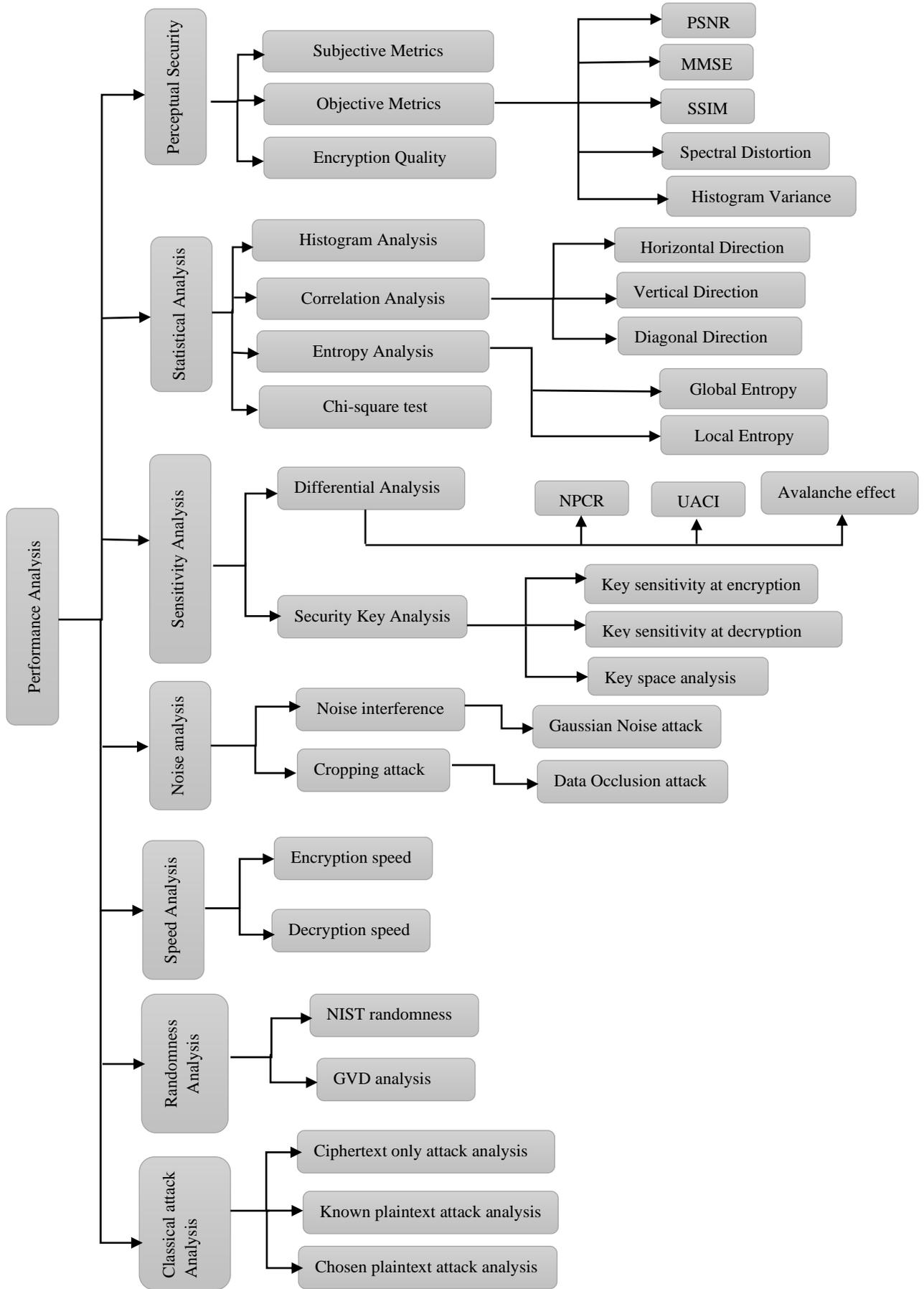

**Fig. 4**. Performance requirements of image encryption scheme

*3.1 Perceptual Security analysis*

Perceptual security can be investigated with some subjective metrics [180]. The ciphertext can be classified into typical quality levels as shown in Table 6. QL0: signifies a completely recognizable image which indicates that the encryption is not valid, QL1: signifies a partially recognizable image contour like edges and boundaries are visible but the texture is not clear. QL2: signifies that the image is completely unintelligible and is considered perceptually secure.

**Table 6** Subjective metrics for perceptual security analysis

| Quality Level | Ciphertext quality |
|---|---|
| QL0 | Image contours are completely recognizable |
| QL1 | Partially recognizable contours of the image |
| QL2 | Completely unintelligent/ white noise like image |

Another measure of perceptual quality is done by evaluating a set of parameters for comparison of encrypted images with reference to the plain image. Some of the commonly used objective metrics are explained below:

*Peak signal to noise ratio (PSNR)*: PSNR is the measure of spectral information in an image. A higher value indicates greater similarity in the test images. In an encryption algorithm, PSNR values are evaluated to quantify the dissimilarity in the encrypted image with respect to plain image. During decryption, the same measure indicates the efficacy of the algorithm in the reverse process. Practically $PSNR \geq 28$ indicates that the test images are similar. For any pair of images, plain image ($P$) and ciphered image ($C$), the $PSNR$ is mathematically defined as:

$$PSNR(P,C) = 10 log_{10} \frac{(L-1)^2}{\frac{1}{MN}\sum_{i=1}^{M}\sum_{j=1}^{N}[P_{i,j} - C_{i,j}]^2} \quad (9)$$

where $L$ represents the highest intensity level which is $2^8$ for an 8-bit image. $M \times N$ represents the size of the image $P_{i,j}, C_{i,j}$ are $(i,j)th$ pixel values of the plain and ciphered image.

i. *Mean square error (MSE)*: It is also an error metric like PSNR that indicates the dissimilarity between the test images. In an ideal case, for two similar images, $MSE$ should be zero. $PSNR$ and $MSE$ are mathematically related to each other as:

$$PSNR(P,C) = 10 log_{10} \frac{(L-1)^2}{MSE} \quad (10)$$

$$\therefore MSE = \frac{1}{MN}\sum_{i=1}^{M}\sum_{j=1}^{N}[P_{i,j} - C_{i,j}]^2 \quad (11)$$

ii. *Spectral Distortion measure (SD)*: It indicates the spectral dissimilarity between the reference image and test image. The SD measure evaluates as to how far is the spectrum of the test image from that of the reference image. The spectral distortion is defined as:

$$SD(P,C) = \frac{1}{MN}\sum_{u=1}^{M}\sum_{v=1}^{N}|F_P(u,v) - F_C(u,v)| \quad (12)$$

where $F_P(u,v), F_C(u,v)$ are Fourier transforms of plain image, $f_P(m,n)$ and encrypted image, $f_C(m,n)$ respectively.

iii. *Structural Similarity Index Measure (SSIM)*: Wang et al. [181] proposed a metric based on the human visual system (HVS) that considers biological factors viz. luminance, contrast and structural comparison between the image and a reference image. This measure known as *SSIM*, is used to quantify the visual image quality.

$$SSIM(x,y) = f(l(x,y), c(x,y), s(x,y)) \quad (13)$$

where *l(x,y)*, *c(x,y) and s(x,y)* are luminance, contrast and structural comparison respectively. For any two pairs of images $P$ and $C$, it is mathematically defined as:

$$SSIM(P,C) = \frac{(2\mu_P\mu_C + C_1)(2\sigma_{PC} + C_2)}{(\mu_P^2 + \mu_C^2 + C_1)(\sigma_P^2 + \sigma_C^2 + C_2)} \qquad (14)$$

where $C_1$ and $C_2$ are two constants, $\mu$, $\sigma$ indicate mean and standard deviation respectively. *SSIM* is '1' for exactly similar images and '-1' for perfectly dissimilar images. Therefore $SSIM \in [-1,1]$.

iv. *Histogram variance*: In order to quantify the uniformity of cipher images, variances of histograms are evaluated [182]. Variances are also evaluated for two different cipher images that are encrypted from two different secret keys on the same plain images. The lower values of variance indicate higher uniformity. The variance of histogram is mathematically evaluated as:

$$var(Z) = 1/n^2 \sum_{i=1}^{n}\sum_{j=1}^{n} \frac{1}{2}(z_i - z_j)^2 \qquad (15)$$

where $Z = \{z_1, z_2, z_3, \ldots z_{256}\}$ is vector of histogram values, $z_i, z_j$ are the number of pixels that have grey values equal to *i* and *j* respectively.

v. *Encryption Quality* is a subjective measure that collectively evaluates an algorithm for the level of security it provides. There are 4 different levels for evaluation as explained in Table 7.

**Table 7** Evaluation of encryption quality

| Security Level | Performance |
|---|---|
| SL0 | High cryptography security + High perceptual equality (QL2) |
| SL1 | High cryptography security + Low perceptual security (QL0, QL1) |
| SL2 | Low cryptography security + High perceptual security (QL2) |
| SL3 | Low cryptography security + Low perceptual security (QL0, QL1) |

## 3.2 Statistical analysis

According to Shannon's communication theory of perfect secrecy [183], "It is possible to evaluate most of the encryption techniques by statistical analysis". He suggested two methods for such analysis. One is histogram analysis and another is correlation analysis for the adjacent pixels in the encrypted image.

*Histogram Analysis:* Histogram is the pixel frequency distribution where each grey level is plotted for the number of pixels with that particular value in the image. An effective cryptosystem should be able to generate ciphertext with fairly uniform histograms which are also significantly different from the plaintext.

*Chi-square test*: In order to verify the uniformity of the histogram, a chi-square test is performed [184] defined as:

$$\chi_{test}^2 = \sum_{k=1}^{K} \frac{(o_i - e_i)^2}{e_i} \qquad (16)$$

where *k* is gray-level (256 for 8-bit image), $o_i, e_i$ are the observed and expected times occurrence of each gray-level, respectively. The test is performed with different significance levels (generally at 0.05) for a null hypothesis.

*Correlation Analysis*: For a perceptually meaningful image, the correlation between adjacent pixels is very high. It is necessary for an effective cryptosystem to significantly reduce these correlation values by decorrelating them in the encrypted domain. For such analysis, either all or a few pixels are randomly selected and correlation plots are obtained for horizontally, vertically and diagonally adjacent pixels. The correlation plots in each direction should display the pixels to be uniformly scattered over the entire intensity range. For quantitative analysis, correlation coefficients are evaluated for two adjacent pixels in horizontal, vertical and diagonal directions by using Eqs. (17)-(19). For $x_i$, $y_i$ as gray values of ith pair of selected adjacent pixels,

$$\rho_{(x,y)} = \frac{cov(x,y)}{\sqrt{D(x)}\sqrt{D(y)}} \qquad (17)$$

where $cov(x,y) = E[x - E(x))(y - E(y))]$

$$= \frac{1}{N}\sum_{i=1}^{i=N}[(x_i - \frac{1}{N}\sum_{i=1}^{i=N}x_i) * (y_i - \frac{1}{N}\sum_{i=1}^{i=N}y_i)] \tag{18}$$

$$D(x) = \frac{1}{N}\sum_{i=1}^{i=N}(x_i - \frac{1}{N}\sum_{i=1}^{i=N}x_i)^2, \quad D(y) = \frac{1}{N}\sum_{i=1}^{i=N}(y_i - \frac{1}{N}\sum_{i=1}^{i=N}y_i)^2 \tag{19}$$

For a perfect similar image, the correlation coefficient is unity and is -1 for completely dissimilar images. Thus $\rho_{(x,y)} \in [-1,1]$.

*Entropy Analysis*: Information entropy is a mathematical property that depicts the randomness associated with the information source. The entropy of a message source *s* is given as:

$$H(d) = -\sum_{i=0}^{L-1} P(s_i)log_2 P(s_i) \tag{20}$$

where *L* is the highest intensity value of pixels in image, $s_i$ is the *ith* symbol in message, P(.) refers to the probability. The entropy defined in Eq. (20) is termed as Shannon's entropy [183]. Besides, a local entropy has been recently proposed [185] as an extension of Shannon's entropy measure. It is the mean entropy of several randomly selected non-overlapping blocks of information source. For an 8-bit image, *L=256*, there are *K=30*, nonoverlapping blocks to be randomly selected from the image with each block having 1936 pixels ($T_B$=1936). Therefore this entropy measure is also termed as ($K,T_B$)-Local entropy and is evaluated using Eq.(21)

$$\overline{H_{k,T_B}}(S) = \sum_{i=1}^{k}\frac{H(S_i)}{k} \tag{21}$$

where $S_i$ are randomly selected non-overlapping image blocks with $T_B$ pixels in each block of *S* with total of *L* intensity scales.

## 3.3 Sensitivity Analysis

*Key Sensitivity analysis:* The sensitivity of an encryption scheme can be evaluated in two aspects: 1) at encryption stage which means that a completely different ciphertext should be generated with a very minute change in the input key value, 2) at the decryption stage, the ciphertext should not be correctly recovered if there is very slight change in the correct key values. Key sensitivity (*KS*) is mathematically defined as:

$$KS = \frac{1}{M \times N}\sum_{m=1}^{M}\sum_{n=1}^{N} C_1(m,n) \otimes C_2(m,n) \times 100\% \tag{22}$$

where $C_1$ and $C_2$ are two different ciphered images with slight change in key values corresponding to same plain image, *P*. $M \times N$ is total number of image pixels in the image.

$$C_1(m,n) \otimes C_2(m,n) = \begin{cases} 1, C_1(m,n) \neq C_2(m,n) \\ 0, C_1(m,n) = C_2(m,n) \end{cases} \tag{23}$$

The value of KS should be as close to 100% [180].

*Key space analysis:* Key space refers to the set of all possible keys that are used in encryption of information. A brute force attack is possible if an intruder manages to make an exhaustive search on the set of possibilities until the correct one is found. Thus, feasibility of brute-force attack depends on the total number of valid keys. This number is an important feature to determine the strength of a cryptosystem and it has to be large enough ($> 2^{100}$) [119] as per today's computing power.

*Differential analysis:* With reference to plaintext, the sensitivity refers to change in ciphertext with slight change in plaintext. This is termed as differential analysis where an adversary can change a single pixel in plaintext and compare the corresponding ciphertexts to get some clue about secret keys. The diffusion property of a cryptosystem enables it to spread any change in plaintext to the entire ciphertext. There are two indicators for numerical evaluation of resistance to such attack: NPCR (number of pixel change rate) and UACI (unified average change in intensity). Theoretically, the closer values of *NPCR* and *UACI* are 99.6093% and 33.4635% respectively indicating the effectiveness of the applied algorithm [186]. These indicators are mathematically defined as:

$$NPCR = \frac{1}{M \times N} \sum_{i,j} D(i,j) \times 100\% \qquad (24)$$

$$UACI = \frac{1}{M \times N} \sum_{i,j} \frac{|C(i,j) - \widetilde{C}(i,j)|}{L-1} \times 100\% \qquad (25)$$

where $C$, $\widetilde{C}$ are two encrypted images with the same keys but with a slight change in the corresponding plain-image of size, $[M\ N]$ with the highest intensity value, $L$.

$$D(i,j) = \begin{cases} 1, & C(i,j) \neq \widetilde{C}(i,j) \\ 0, & otherwise \end{cases} \qquad (26)$$

*Avalanche effect:* The avalanche criterion is referred to as an average number of bits that differ between $C$ and $\widetilde{C}$ while changing a pixel in plaintext. The ideal value of the avalanche effect is 0.5 (50%).

### 3.4 Noise analysis

The communication channels over which the image information is transferred are responsible for the addition of some noise in the form of degradation or distortion. The performance of a cryptosystem in such a scenario requires analysis. Gaussian noise with zero mean and varying values for variance is added to the encrypted image for *Gaussian noise analysis*. The quality of the decrypted image is checked in perceptual as well as numerical terms with different variances in noise [112, 187]. The results thus obtained, are compared for the noise analysis. The *Occlusion attack* refers to the loss of data or cropping of a portion of the image due to noisy channels. The cryptosystem should be capable of recovering the appropriate amount of information even after some occlusion in data. In order to check for the robustness to occlusion attack, some pixels of encrypted image (10%,15%,25%,50%,75%) are cropped and corresponding decrypted image quality is evaluated in perceptual and numerical analysis [99] [50] [187].

### 3.5 Speed analysis

Speed analysis refers to the critical execution time for forward and reverse process in an encryption scheme. As typical configuration and capacity of a system greatly determines its computation speed, therefore a comparison of encryption and decryption time is a trivial task. Different machines perform differently. However, time analysis is an important feature, especially where real-time application is involved. Time analysis is done in terms of encryption time and decryption time separately. Generally, a large sample set of images are considered for evaluating the average time taken in the encryption and decryption process on a present-day commonly used system configuration.

### 3.6 Randomness analysis

*NIST SP800-22* is a statistical test suite for random and pseudorandom number generators that are used for cryptographic applications. The advantage of this test suite is that it does not require any assumptions on the generator. Rather, it only looks for a particular statistical recurrence in the generated sequence (random). It consists of 15 *p*-value based tests that include frequency test, run test and spectral test. These tests are generally not used in transform-based cryptography. However, we mention it here due to usage of it in some classical methods of image encryption.

*GVD analysis*

The gray value difference of a pixel form its four neighbouring pixels in an image is given by:

$$G(i,j) = \sum \frac{[I(i,j) - I(i',j')]}{4} \quad (27)$$

The average difference in gray values corresponding to each pixel in image is

$$G_{av}(i,j) = \frac{1}{(M-2)(N-2)} \sum_{i=2}^{M-1} \sum_{j=2}^{N-1} G(i,j) \quad (28)$$

Thus, Gray value difference (GVD) parameter [188] of an encryption scheme is defined as:

$$GVD = \frac{G_{av}^P(i,j) - G_{av}^C(i,j)}{G_{av}^P(i,j) + G_{av}^C(i,j)} \quad (29)$$

where $G_{av}^P$ and $G_{av}^C$ are the average differences in gray values for original plain image and ciphered image respectively. The ideal value of GVD parameter is unity. For a good encryption scheme, this parameter should be as close to 1.

*3.7 Classical attack analysis*

In cryptography, classical attacks are launched to cryptanalyze an encryption scheme. The adversary can have certain information regarding plain text or ciphertext that provide for cryptanalysis. If the adversary has access to set of ciphertext then it can launch a *ciphertext only attack*. If it is able to get access to set of plain texts and corresponding ciphertexts then a *known plaintext attack* can be launched. In a *chosen plaintext attack,* it is assumed that the adversary has access to arbitrary plaintexts and can obtain the corresponding ciphertexts. From the above-stated assumptions, a chosen plaintext attack provides the most information to the adversary. Thus, if a cryptosystem is able to resist chosen plaintext attack, it is believed to be able to resist other classical attacks as well [189] [133]. Therefore, an image encryption scheme should have excellent diffusion properties for providing robustness to a chosen plaintext attack analysis.

## 4. Comparative analysis

As shown in Table 8, each of the proposed schemes is accompanied by the parameters used to evaluate the encryption algorithm and the technique that is merged with the fractional transform. We have categorized these techniques into eight, as reality preserving (T01), chaos theory based (T02), compressive sensing (T03), multiple parameters (T04), DNA sequence (T05), cellular automata (T06), double image encryption (T07.1), multiple image encryption (T07.2), with watermarking (T08). The comparative analysis is based on the results available for *Lena* image only. Table 9 illustrates the subjective comparison for the same references as listed in Table 8 along with the probable vulnerabilities associated with each of them. These vulnerabilities are expressed as V01-V07 (mentioned below the Table 9). It is worth mentioning here that the vulnerabilities of each scheme can be removed by specific methodology in practice.

**Table 8** Comparative analysis for performance metrics of proposed schemes (for *Lena* image)

| [Reference] | Year | Technique used | Correlation analysis | | | Average Entropy | Key space | Average NPCR(%) | Average UACI(%) | Encryption Quality |
|---|---|---|---|---|---|---|---|---|---|---|
| | | | Horizontal | Vertical | Diagonal | | | | | |
| Ref. [116] | 2021 | T02, T05 | 0.0015 | 0.0014 | 0.0059 | 7.9952 | $10^{247}$ | 99.6348 | 33.5816 | SL0 |
| Ref. [147] | 2021 | T02, T03,T04 | -- | -- | -- | -- | $2^{259}$ | -- | -- | SL1 |
| Ref. [117] | 2021 | T02, T05 | 0.0033 | −0.0099 | −0.0046 | 7.9768 | $10^{228}$ | 99.5956 | 33.8798 | SL1 |
| Ref. [102] | 2020 | T02, T05 | 0.0693 | 0.0610 | −0.0242 | 7.9991 | --- | 99.5677 | 33.4353 | SL0 |
| Ref. [172] | 2020 | T02,T07.2 | 0.0223 | 0.0187 | 0.0137 | 1.0149 | $10^{70}$ | 99.4664 | 34.1316 | SL1 |
| Ref. [187] | 2020 | T01,T02 | −0.0006 | −0.0057 | 0.0009 | 7.9938 | $10^{102}$ | 99.6006 | 34.6379 | SL0 |
| Ref. [115] | 2019 | T01,T02, T07.2 | 0.0036 | −0.0038 | 0.0023 | 7.99 | -- | -- | -- | SL2 |
| Ref. [52] | 2018 | T02,T07.1 | 0.0001 | −0.0029 | −0.0019 | 7.5907 | -- | 99.7400 | 0 | SL2 |
| Ref. [143] | 2018 | T02, T03 | 0.0127 | 0.0101 | 0.0139 | -- | $10^{136}$ | -- | -- | SL2 |

| Ref. [114] | 2018 | T01, T02, T04 | −0.0001 | −0.0014 | 0.0004 | -- | -- | 99.8640 | 33.3330 | SL0 |
| Ref. [112] | 2018 | T01, T02, T04 | 0.0015 | 0.0017 | −0.0033 | -- | $10^{98} = 2^{325}$ | 99.9949 | 33.3616 | SL0 |
| Ref. [103] | 2018 | T02 | 0.0020 | −0.0007 | 0.00006 | 7.4739 | -- | -- | -- | SL0 |
| Ref. [104] | 2018 | T01, T02, T07.1 | -- | -- | -- | | | | | SL3 |
| Ref. [108] | 2017 | T02 | 0.01513 | −0.0024 | −0.0045 | 7.9974 | $2^{297}$ | -- | -- | SL2 |
| Ref. [137] | 2017 | T02 | 0.1068 | 0.0766 | 0.0182 | -- | $\approx 10^{16}$ | -- | -- | SL3 |
| Ref. [144] | 2017 | T02, T03 | 0.0909 | 0.2389 | 0.0126 | -- | $10^{37}$ | -- | -- | SL2 |
| Ref. [171] | 2017 | T07.2 | 0.0249 | 0.0505 | 0.0280 | -- | $27^5 \times 30^5$ | 99.6279 | 33.4599 | SL2 |
| Ref. [169] | 2016 | T02, T07.1 | -- | -- | -- | -- | $10^{55}$ | -- | -- | SL2 |
| Ref. [100] | 2015 | T02 | 0.0104 | 0.0299 | 0.0062 | -- | $10^{34} \times 13^5 \times 11^5$ | -- | -- | SL2 |
| Ref. [142] | 2015 | T02, T03 | 0.0119 | 0.0925 | 0.0325 | -- | $10^{64}$ | -- | -- | SL2 |
| Ref. [54] | 2015 | T07.1 | 0.0093 | 0.0172 | 0.0021 | -- | -- | -- | -- | SL2 |
| Ref. [170] | 2014 | T07.2 | 0.0040 | −0.0018 | 0.0266 | 7.9976 | -- | -- | -- | SL2 |

It is evident from the values in Table 8 that studies in which chaos-based permutation or substitution is merged with fractional transform domain have higher entropy measure, low correlation coefficients, high NPCR and UACI, higher key space, excellent key sensitivity, robustness to noise and data occlusion attacks, hence having higher security levels. Reality preserving algorithm has contributed toward the digital implementation of optical transforms and has enabled researchers to overcome major limitations regarding complexity issues of fractional transforms in the digital domain. Compressive sensing is used to reduce the data deluge while dealing with large images for encryption but their performance is marginal in terms of higher correlation coefficients and vulnerability to leakage in information.

CS-based encryption schemes are highly complex [190] and reconstruction is time-consuming. It has been observed in the results of the above-reviewed articles that CS-based schemes lack uniform histograms in the encrypted domain and CC values are considerably higher. Also, CS-based simultaneous compression and encryption schemes are vulnerable to cryptanalysis due to linearity [191]. In a broad sense, if the plaintext is sparse, the key of the cryptosystem may not be safe as it is possible to exploit the prior sparsity knowledge to extract information of the key from ciphertext. The key and the plaintext may be partly accessed by using some information processing technology such as Blind source separation (BSS) [192].

**Table 9**. Comparative analysis for subjective parameters ( Refer Table. 8 for performance metrics)

| Reference | Metrics for perceptual analysis | Noise analysis | Occlusion attack | Classical attacks | Differential attack | Statistical attack | Time analysis | Probable Vulnerabilities |
|---|---|---|---|---|---|---|---|---|
| Ref. [116] | ✓ | ✓ | ✓ | ✓ | ✓ | ✓ | ✓ | V06, V07 |
| Ref. [147] | ✓ | ✓ | ✓ | ✗ | ✗ | ✓ | ✗ | V02, V05, V07 |
| Ref. [117] | ✓ | ✓ | ✓ | ✓ | ✓ | ✓ | ✓ | V07 |
| Ref. [102] | ✗ | ✗ | ✗ | ✗ | ✓ | ✓ | ✗ | V01, V03, V06, V07 |
| Ref. [172] | ✓ | ✓ | ✓ | ✓ | ✓ | ✓ | ✗ | V03, V07 |
| Ref. [187] | ✓ | ✓ | ✓ | ✓ | ✓ | ✓ | ✗ | V03, V07 |
| Ref. [115] | ✓ | ✗ | ✗ | ✗ | ✗ | ✓ | ✗ | V01, V02, |
| Ref. [52] | ✗ | ✓ | ✓ | ✓ | ✓ | ✓ | ✓ | V02 |
| Ref. [143] | ✗ | ✗ | ✗ | ✓ | ✗ | ✓ | ✗ | V01, V02, V07 |
| Ref. [114] | ✓ | ✓ | ✓ | ✓ | ✓ | ✓ | ✗ | V03, V07 |
| Ref. [112] | ✓ | ✓ | ✓ | ✓ | ✓ | ✓ | ✗ | V07 |
| Ref. [103] | ✓ | ✗ | ✗ | ✗ | ✗ | ✓ | ✗ | V03, V04, V05, V07 |
| Ref. [104] | ✓ | ✗ | ✗ | ✗ | ✗ | ✓ | ✗ | V02, V04, V06, V07 |
| Ref. [108] | ✓ | ✗ | ✗ | ✗ | ✗ | ✓ | ✗ | V02, V03, V07 |
| Ref. [137] | ✗ | ✓ | ✓ | ✗ | ✗ | ✓ | ✗ | V02, V03, V07 |

| Ref. [144] | ✓ | ✓ | ✓ | ✗ | ✗ | ✓ | ✗ | V01, V02, V07 |
| Ref. [171] | ✗ | ✓ | ✓ | ✗ | ✓ | ✓ | ✓ | V01 |
| Ref. [169] | ✗ | ✓ | ✓ | ✗ | ✗ | ✓ | ✗ | V02, V03, V07 |
| Ref. [100] | ✗ | ✓ | ✓ | ✓ | ✗ | ✓ | ✗ | V02, V07 |
| Ref. [142] | ✓ | ✓ | ✓ | ✗ | ✗ | ✓ | ✓ | V02, V06 |
| Ref. [54] | ✓ | ✓ | ✓ | ✗ | ✗ | ✓ | ✗ | V02, V07 |
| Ref. [170] | ✓ | ✓ | ✓ | ✓ | ✗ | ✓ | ✗ | V02, V07 |

**V01: High complexity, V02: Low encryption quality, V03: Dependent on diffusion, V04: Smaller key space, V05: poor efficiency, V06: Lossy, V07: may not be applicable for real time applications**

Multiple parameter-based fractional transform schemes perform better than fixed/single transform order based schemes. This is due to enlarged key space and better uniformity in encrypted histograms. However, there are some deficiencies related to multiple parameter schemes [193], [194], [195] due to linearity that need to be avoided. The linear relation among consecutive transform orders and periodicity is the major limitation that can lead to multiple decryption keys corresponding to an encryption key. This depicts its vulnerability to various attacks. To overcome this issue, it is necessary to introduce some means of breaking the linear relationship among consecutive transform orders or by careful selection of transform orders through a random selection scheme [152], [187].

DNA sequence operation is little less explored with optical transforms. However, it is able to enhance security with increased key space and randomness in encrypted data. Double and multiple image encryption schemes are preferred for speed and increasing encryption efficiency. Watermarking is another domain where fractional transforms are used to encrypt the watermark before embedding in a blind watermark scenario. The encryption of the watermark logo in the collective time-frequency domain increases the robustness to various attacks.

## 5. Observations based on published literature

In an exhaustive search performed in the month of December 2021 on the various online databases: ACM Digital library, Elsevier, Google Scholar, IEEE explore, Springer link, Taylor and Francis and Wiley for the number of research papers published related to the encryption of different multimedia contents during the period 2015-2021. The pictorial view to highlight the percentage of papers published on the encryption of various multimedia contents like: images, video, audio, text data etc. has been shown in Fig. 5.

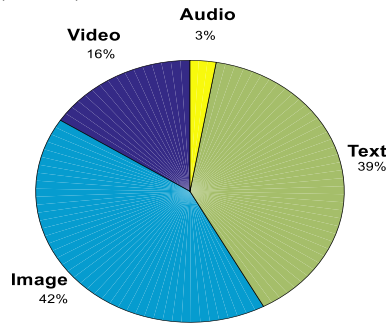

**Fig. 5.** Percentage of research papers published on the encryption of various multimedia contents like: images, video, audio, text data during 2015-2021.

According to search results, it is observed that the number of publications is majorly in text and image encryption. However, the number of image encryption works is dominating with 42% of all the metadata available. We believe, it is due to the wide application area of image data, from platforms like social media to sensitive data like military and telemedicine fields. Almost every sector of communication is dependent on image transmission in one way or the other. It is also observed that amongst various mathematical implementations of the fractional transforms, fractional Fourier transform (FrFT) is most popular with more than 60% of the total publications in fractional integral based image encryption schemes. This is followed by fractional wavelet transform (FrWT) with a contribution of 16%, fractional Hartley transform, FrHT (10%), fractional Cosine transform, FrCT (7%) and the remaining few on other transforms (viz. Mellin, angular, sine etc.).

As the present manuscript is mainly concerned with image encryption using optical/fractional integral transforms, therefore, we narrowed down our search for the number of papers published year-wise on the fractional transform-based image encryption schemes. Fig.6 illustrates a graphical representation of the related publications in all the major online databases during the period 2015-2021.

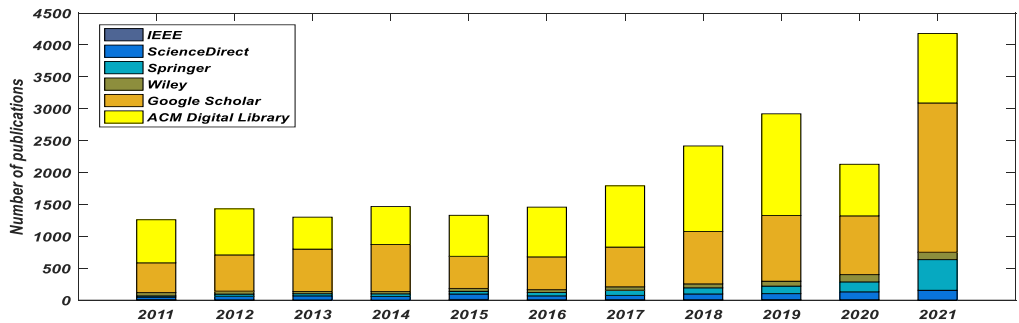

**Fig. 6.** Number of papers published on fractional integral transform-based image encryption schemes on various online databases

It is observed that the number of publications on image encryption in the fractional transform domain has considerably increased every year. This gives testimony to the fact that, with the advent of evolutionary algorithms based on fractional integral transforms in the digital domain has increased its popularity and is receiving significant attention from the researcher community.

It has been also observed that most of the encryption algorithms with fractional transform as the main component are evaluated for statistical analysis, noise attack and occlusion attack analysis only. This is probably the reason for less popularity of optical transform-based image encryption schemes as compared to purely chaos-based schemes or other number theory-based approaches. According to a recent survey on color image encryption [65], only 8.65% of the proposed schemes are based on optical transforms. In order to widen the contribution of optical transform-based schemes to image encryption, certain limitations need solutions for encouraging practical implementations.

In Section 2.3, we have described the categorization of fractional transform-based image encryption schemes in accordance with the strategical amalgamation of the fractional transform domain with other evolutionary methods. There are total of eight major categories T01 to T08 (one of them T07 having two subcategories). In Fig.7, we have shown the relative contributions in terms of the number of papers published in each of these categories so far. We observe that the major contributions come from the T07: Double Image/Multiple Image category, followed by T02: chaos-based, T08: Watermarking, T03: Compressive Sensing, T01: Reality Preserving category T04: Multiple/fixed parameter transforms, T05: DNA Sequences, and least in T06: Cellular Automata.

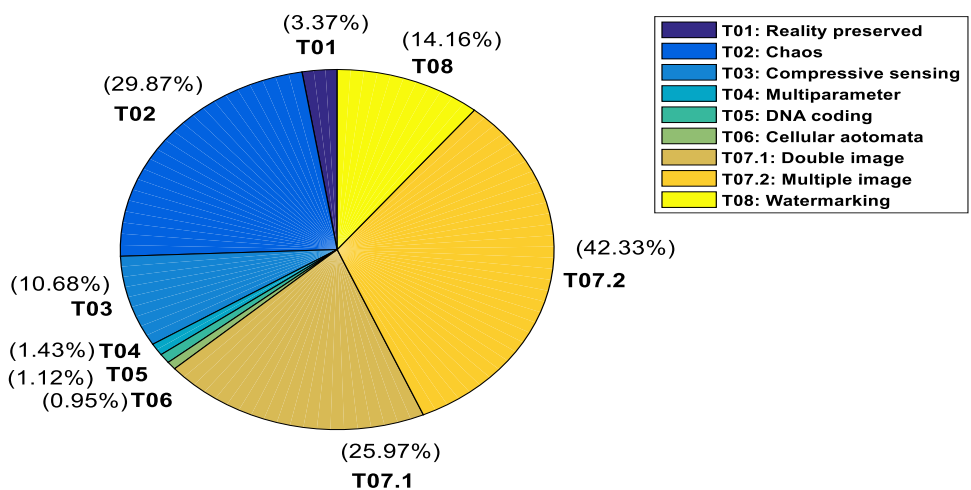

**Fig. 7.** Relative contribution in terms of the number of papers published belonging to different categories (T01-T08) of fractional transform-based image encryption techniques

Based on the observations related to security levels and vulnerabilities mentioned in Table 8 and Table 9, we elaborate on the possible ways to overcome some limitations. Most of the algorithms mainly lack in the following aspects: (1) uniform histograms, (2) entropy measure, (3) smaller key space, (4) differential analysis, (5) classical attack analysis, (6) speed analysis. In the discussion below, we try to highlight some of the possible solutions as:

- *Uniform Histogram:* A majority of fractional transform-based image encryption schemes produce cipher images having Gaussian distribution like histograms [51] [114] [104]. It is due to the fact that the energy of a transform is concentrated at the centre. Authors have claimed the robustness of encryption schemes only on the basis of similarity in the distribution of histograms irrespective of the content of the plain image. The entropy measure for such distributions has values that is significantly less than the ideal value (8 for 256 intensity levels image). However, in cryptography, it is expected that the cipher image pixels should have a uniform distribution over the entire intensity ranges having entropy measure very near or equal to the ideal value. This points to some information leakage, that can make a scheme vulnerable to entropy attacks. To overcome such limitation, a hybrid algorithm in which fractional integral transform domains are amalgamated with chaos based pseudorandom substitutions should be used.
- *Smaller Key space:* Adopting multiple layer security for image encryption algorithm will lead to an increase in key space. Apart from this, making a selection of transform orders to depend on some chaotic parameters or a similar analogy will result in larger key space [114] [187]. Most of the proposed schemes have added a permutation layer along with the transform domain. Some of the schemes that are based on permutation and substitution paradigm are able to offer larger key space to overcome brute force attack.
- *Differential Analysis:* In order to fulfil the requirement of effective encryption algorithm, the scheme should be able to resist differential attack analysis. The parameters NPCR and UACI are its measures. From Table 9, it is clear that majority of schemes lack such analysis. Even if done, the UACI values are not optimum or even *zero*. This is due to the fact that there is no significant change in intensity values with a single pixel change in input. Therefore, for a successful strategy, the change should be diffused over the entire image coefficients. One of the solutions to this issue is to make the initial parameters of diffusion scheme to depend on some significant feature of the input image like mean or average values.
- *Time analysis:* A run time for an encryption algorithm refers to the time required for its execution. Various factors need to be considered for time analysis like the size of image, system configuration, programming language etc. [63]. To compare the computational performance of an algorithm, is a crucial task as different host machines have their own set of configurations. Due to this reason, some researchers have used an average time Vs size paradigm to evaluate computational performance [120] wherein input images with variable size are selected and the average time of encryption is evaluated by using large set of different keys. Fractional transform-based encryption schemes have inherent advantage of high speed and parallel processing. However, while merging of these schemes with other domains like chaos etc., computational optimization should be taken care of. In summary, there should be trade-off management between complexity and security while designing an algorithm and some optimum suggestion for the choice of parameters, number of rounds etc. should be given.
- *Careful Selection of chaotic maps:* The chaotic maps wherever used in an encryption scheme, need a careful selection. As most of the schemes that are reviewed have employed one dimensional chaotic map [50] [103] [129]. Although 1D maps are simplest in hardware implementation but are less secure. For instance, 1D logistic maps have some periodic windows in the chaotic range [196] and that Arnold transform also has periodicity [197], hence are vulnerable. At the same time, the higher dimensional chaotic maps are sometimes secure but complex. To keep a balance, it is recommended to use a coupled map scheme where two or more 1D chaotic maps are coupled for enhanced security [52] and also robust chaotic maps may be used with proper specification of the range of parameters where robust chaos is observed. Prior to selection of such chaotic map, a proper bifurcation analysis and investigation of dynamical behaviour in the entire parameter space must be done to identify the suitable regions of parameter space exhibiting robust chaos.

## 6. Conclusion

The evolution of digital media over the past two decades has revolutionized the development of strategies pertaining to security preservation of the multimedia contents. Encryption is the most effective way to secure the data. It has been observed in the study that out of all the data types, (audio, video, text, image) image data is most frequently used to convey the information. Consequently, the percentage of published work on image encryption is dominating with 42% of all the metadata available. However, cryptography for image data is challenging when it comes to classical methods of encryption due to huge volume of data and also due to the high correlation among adjacent pixel values. Various research works have been proposed in the literature that are specifically suitable for image encryption. Application of fractional integral transforms in image encryption has been an active research area and the review work in this paper is also focussed on the same . The fractional integral transform provide an extra degree of freedom to the encrypted data as the fractional order of the transform is used as secret key.

The aim of this review is to build an understanding of the reader towards application of fractional itegral transforms in image encryption. The initial description of the paper gives a conceptual idea on using these transforms and also the domain-based taxonomy to classify various existing schemes in the literature. The optical image encryption that comprises of optical setup and double random phase encoding (DRPE) has been discussed. Few recent review works and cryptanalysis of these schemes are tabulated and analyzed. The digital implementation of the fractional integral transforms is discussed with its analogy to the optical setup. Further, various algorithms are categorized in accordance with their merging techniques and a comprehensive review is presented on some of the most recently published articles. The performance criteria and standards to be followed have been discussed. A performance comparison in tabular format is presented for objective as well as subjective metrics of some of the recent publications. Finally, based on the observations, some major concerns are listed and a few constructive guidelines are provided. This work intends to provide the readers with an understanding of why and how fractional integral transformations are applicable to the encryption of images. In addition, the study highlights some vulnerabilities and threats associated with the usage of fractional transforms along with the probable solutions that may help in the future design and development of hybrid and robust encryption schemes.

**Acknowledgements**

One of us (V.P.) acknowledges the MATRICS Grant (MTR/000203/2018) from SERB, DST, Govt. of India.